\documentclass[twocolumn]{aastex6}
\usepackage{graphicx}
\usepackage{amssymb,amsfonts,amsmath,amstext,amsgen,amsopn,amsxtra,indentfirst,times}

\usepackage{listings}
\usepackage{chemformula}
\usepackage{color}

\begin{document}

\title{HELIOS-K 2.0 Opacity Calculator and Open-source Opacity Database for Exoplanetary Atmospheres}

\author{Simon L. Grimm\altaffilmark{1}}
\author{Matej Malik\altaffilmark{2}}
\author{Daniel Kitzmann\altaffilmark{1}}
\author{Andrea Guzm\'{a}n-Mesa\altaffilmark{1}}
\author{H. Jens Hoeijmakers\altaffilmark{1}}
\author{Chloe Fisher\altaffilmark{1,8}}
\author{Jo\~{a}o M. Mendon\c{c}a\altaffilmark{3}}
\author{Sergey N.Yurchenko\altaffilmark{4}}
\author{Jonathan Tennyson\altaffilmark{4}}
\author{Fabien Alesina\altaffilmark{5}}
\author{Nicolas Buchschacher\altaffilmark{5}}
\author{Julien Burnier\altaffilmark{5}}
\author{Damien Segransan\altaffilmark{5}}
\author{Robert L. Kurucz\altaffilmark{6}}
\author{Kevin Heng\altaffilmark{1,7}}

\altaffiltext{1}{University of Bern, Center for Space and Habitability, Gesellschaftsstrasse 6, CH-3012, Bern, Switzerland.  Emails: simon.grimm@csh.unibe.ch, kevin.heng@csh.unibe.ch}
\altaffiltext{2}{Department of Astronomy, University of Maryland, College Park, MD 20742, USA}
\altaffiltext{3}{National Space Institute, Technical University of Denmark, Elektrovej, DK-2800, Kgs. Lyngby, Denmark}
\altaffiltext{4}{Department of Physics and Astronomy, University College London, London, WC1E~6BT, UK}
\altaffiltext{5}{University of Geneva, Department of Astronomy, CH-1290 Versoix, Switzerland}
\altaffiltext{6}{Center for Astrophysics $\vert$ Harvard \& Smithsonian, Cambridge, MA 02138, USA}
\altaffiltext{7}{University of Warwick, Department of Physics, Astronomy and Astrophysics Group, Coventry CV4 7AL, UK}
\altaffiltext{8}{University of Bern International 2021 PhD Fellowship, Switzerland}

\begin{abstract}
Computing and using opacities is a key part of modeling and interpreting data of exoplanetary atmospheres. Since the underlying spectroscopic line lists are constantly expanding and currently include up to $\sim 10^{10}$--$10^{11}$ transition lines, the opacity calculator codes need to become more powerful. Here we present major upgrades to the \texttt{HELIOS-K} GPU-accelerated opacity calculator and describe the necessary steps to process large line lists within a reasonable amount of time. Besides performance improvements, we include more capabilities and present a toolbox for handling different atomic and molecular data sets, from downloading and preprocessing the data to performing the opacity calculations in a user-friendly way. \texttt{HELIOS-K} supports line lists from ExoMol, HITRAN, HITEMP, NIST, Kurucz, and VALD3. By matching the resolution of 0.1 cm$^{-1}$ and cutting length of 25 cm$^{-1}$ used by the \texttt{ExoCross} code for timing performance (251 s excluding data read-in time), \texttt{HELIOS-K} can process the ExoMol BT2 water line list in 12.5 s. Using a resolution of 0.01 cm$^{-1}$, it takes 45 s, equivalent to about $10^7$ lines s$^{-1}$. 
As a wavenumber resolution of 0.01 cm$^{-1}$ suffices for most exoplanetary atmosphere spectroscopic calculations, 
we adopt this resolution in calculating opacity functions for several hundred atomic and molecular species and make them freely available on the open-access DACE database. 
For the opacity calculations of the database, we use a cutting length of 100 cm$^{-1}$ for molecules and no cutting length for atoms.
Our opacities are available for downloading from \texttt{https://dace.unige.ch/opacityDatabase} and may be visualized using \texttt{https://dace.unige.ch/opacity}.

\end{abstract}

\keywords{planets and satellites: atmospheres}

\section{Introduction}
\label{sect:intro}

Opacities\footnote{Specifically, this term is used to describe specific forms of the cross section per unit mass involving photons/light/radiation and not electron or heavy-particle collisions.} are the cross sections per unit mass of atoms, ions, and molecules. They may be measured in the laboratory or calculated from first principles (by solving the Schr\"{o}dinger equation) and are indispensable in the analysis of spectra across multiple subfields of astronomy and astrophysics.  In the study of exoplanetary atmospheres, opacities are used in a number of fundamental ways: in one-dimensional radiative transfer calculations, three-dimensional general circulation models, and atmospheric retrievals.

Generally, the opacity of an atom, ion, or molecule depends on temperature, pressure, and wavenumber or wavelength.  The number of spectral lines increases exponentially with temperature.  For example, one only needs to compute $\sim 10^5$ lines of water at room temperature (because the weaker lines are not activated) but $\gtrsim 10^9$ lines at $\sim 1000$ K.  Furthermore, since models typically require hundreds, if not thousands, of combinations of temperature and pressure---as well as multiple chemical species---this problem rapidly becomes formidable.

Solving this computational problem requires the judicious design of software and its careful interfacing with hardware, including efficient memory management.  In \cite{gh15}, we took the first major steps toward producing an ultrafast, open-source opacity calculator named \texttt{HELIOS-K} (\texttt{https://github.com/exoclime/HELIOS-K}), which is accelerated by the use of graphics processing units (GPUs).  In the current study, we report major upgrades to \texttt{HELIOS-K} that allow for further order-of-magnitude increases in computational speed.

As a service to the community and to promote scientific reproducibility and transparency, the computed opacities are publicly available on \texttt{https://dace.unige.ch/opacityDatabase}  and \texttt{https://chaldene.unibe.ch/}.  The former offers the possibility of obtaining customized opacities across user-specified arrays of temperature and pressure.  The opacities may be visualized on the Swiss-run DACE database via \texttt{https://dace.unige.ch/opacity}, as we will describe.  We note that the ExoMolOP database provides publicly available opacities as well \citep{Chubb2020}.

\section{Methodology}

The basic methodology of the \texttt{HELIOS-K} opacity calculator was published in \cite{gh15}. The updates on the core methodology include a mostly a new parallelization scheme and faster memory handling, as well as new features of the code. We summarize the supported line profiles in section \ref{profiles} and the background theory in section \ref{background}, describe the difficult aspects of the usage of GPUs in section \ref{sect:GPU}, and report the improvements and updates of \texttt{HELIOS-K} in section \ref{improvments}. In section \ref{sect:prepare}, we describe in detail the handling of the different line list databases. 

\subsection{Line profiles}
\label{profiles}
\begin{figure}[h]
\begin{center}
\includegraphics[width=\columnwidth]{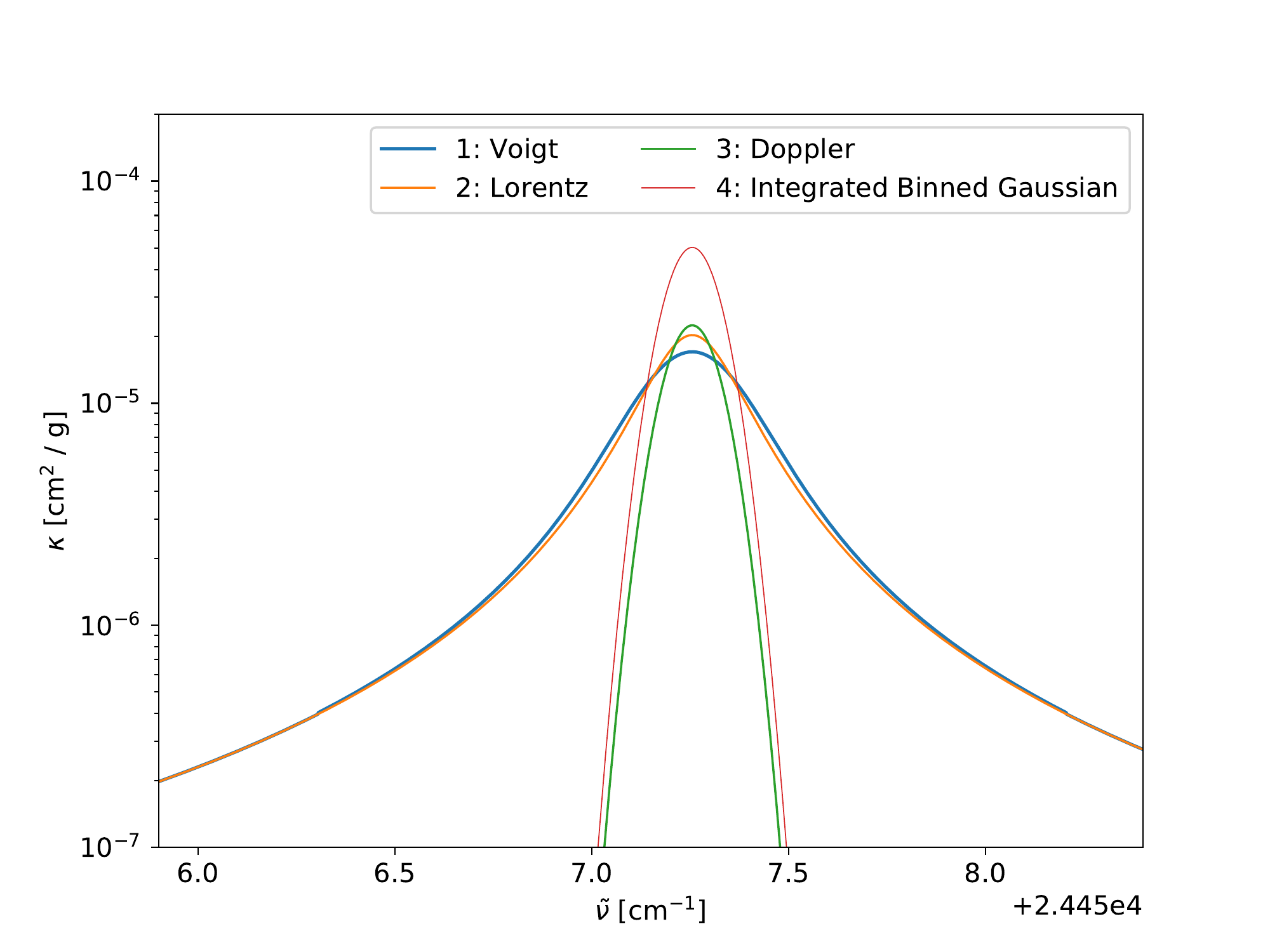}
\end{center}
\caption{Available line profiles from \texttt{HELIOS-K}: Voigt, Lorentzian, Doppler, and the binned Gaussian integrated cross section. The Voigt profile is numerically the most expensive to calculate.}
\label{fig:profiles}
\end{figure}

In most applications, molecular or atomic transition lines are approximated by a Voigt\footnote{It should be noted that name of W. Voigt is usually mispronounced. The correct pronunciation is [fo:kt]; i.e. the vowels are articulated ``oo,'' rather than the commonly used ``oi.''} profile, which is a convolution of a Doppler and a Lorentzian profile. This convolution consists of an infinite integral, which is numerically expensive and requires optimized numerical methods to solve in a reasonable time. Besides the Voigt profile, sometimes pure Doppler and Lorentzian profiles are also used, and for comparison reasons, we also implemented the ``binned Gaussian integrated cross section" used by the \texttt{ExoCross} code \citep{Yurchenko+2018}. \texttt{HELIOS-K} allows one to choose between these four profiles. Figure \ref{fig:profiles} shows an example of the available line profiles.
 
For some species, there are non-Voigt profiles available, e.g. \cite{nonVoigt}, but those are not implemented in the current version of \texttt{HELIOS-K}. However, it is possible to replace individual transition lines with precalculated tables as necessary for certain alkali resonant lines.

\subsubsection{Removing the Plinth}
\begin{figure}[h]
\begin{center}
\includegraphics[width=\columnwidth]{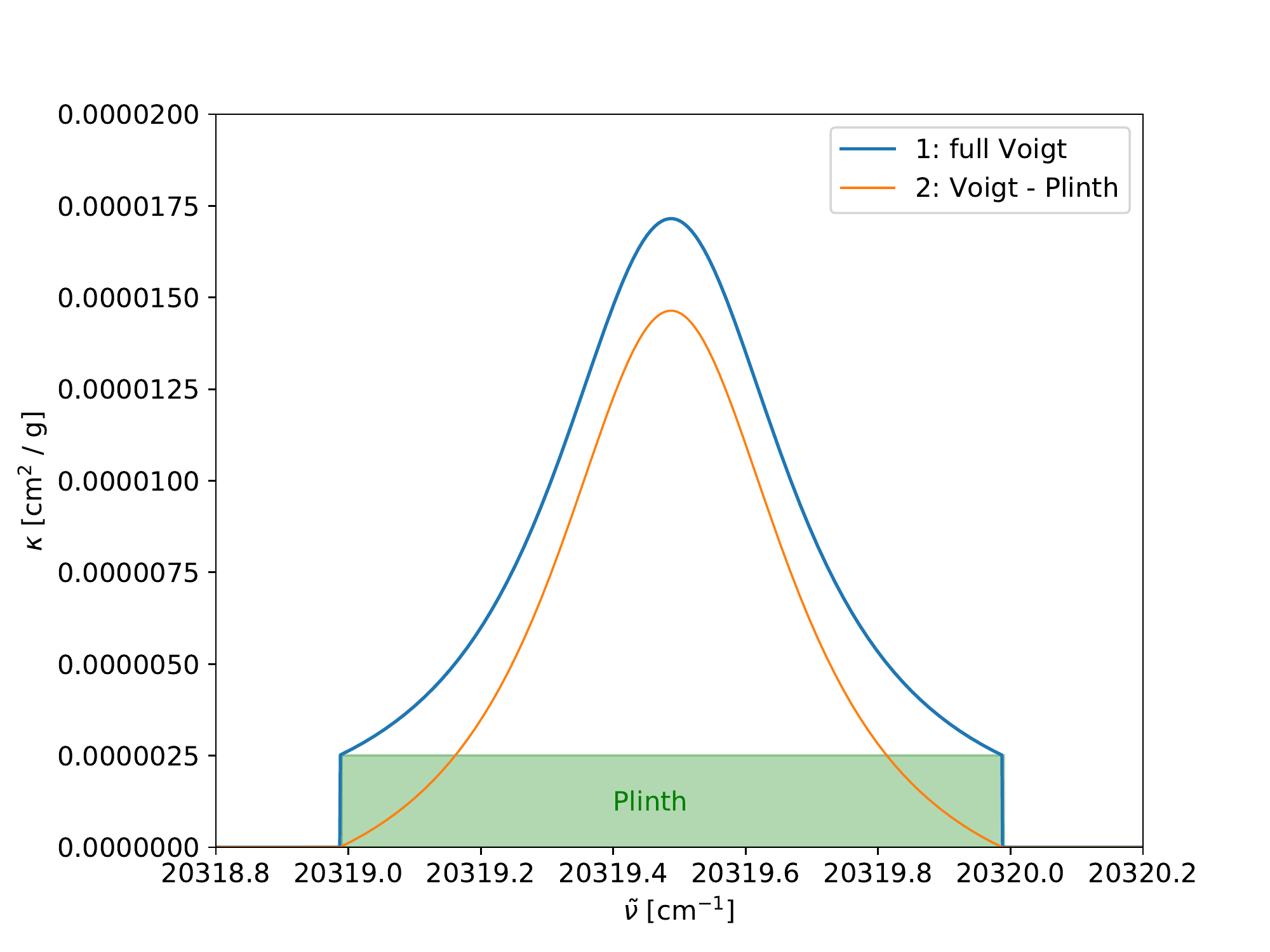}
\end{center}
\caption{When truncated line wing opacities are used in combination with a far wing background, it may be necessary to remove the plinth from the opacities. The height of the plinth is defined as the opacity value at the cutting length position. Shown in blue is a full Voigt profile transition line, and shown in orange is the transition line with the removed plinth. The plinth itself is shown in green.}
\label{fig:plinth}
\end{figure}

\texttt{HELIOS-K} supports the option of removing the plinth (also sometimes called the ``base") from transition lines. The height of the plinth is defined as the opacity value from a transition line at the cutting length position 
and depends on the line intensity and broadening parameters (e.g. \citealt{clough+1989,paynter+2011,Ptashink+2012}). 
Removing the plinth is necessary  when opacities are used in combination with a background opacity that already includes the described plinth for each truncated transition line. This is the case, for example, when using the \texttt{CKD} \citep{clough+1989}, \texttt{MT\_CKD} \citep{Mlawer+2012}, or \texttt{BPS} \citep{paynter+2011} water vapor continuum as an opacity background. Not removing the plinth would lead to double counting of it.
An example of the plinth and how it is removed from a transition line is shown in Figure \ref{fig:plinth}.

\subsection{Background theory}
\label{background}

We follow the definitions of the parameters described in \cite{gh15} and \cite{Heng2017} and repeat only some of them here for a better overview. Molecular or atomic opacities can be approximated by Voigt line profiles, which can be described by two parameters\footnote{In the literature, these parameters are sometimes also called $x$ and $y$}, $a$ and $u$.
These two parameters for the Voigt line profiles are defined as 
\begin{equation}
\label{eq_a}
a = \sqrt{\ln 2} \frac{\Gamma_L}{\Gamma_D}
\end{equation}
and
\begin{equation}
\label{eq_u}
u = \sqrt{\ln 2} \frac{\tilde\nu - \tilde\nu_0}{\Gamma_D},
\end{equation}
where $\tilde{\nu}$ is the wavenumber, $\tilde\nu_0$ is the line center in wavenumber in cm$^{-1}$, and $\Gamma_L$ and $\Gamma_D$ are the Lorentz and Doppler half-widths in cm$^{-1}$, respectively.
The Doppler half-width is given as
\begin{equation}
\label{eq_gammaD}
\Gamma_D = \frac{\tilde\nu_0}{c} \sqrt{\frac{2 \ln2 k_B T}{m}}, 
\end{equation}
where $m$ is the mass in grams (molar mass in \mbox{g mol}$^{-1}$ divided by the Avogadro constant), $k_B=1.3806489 \times 10^{-16}$ cm$^2$ g s$^{-2}$ K is the Boltzmann constant, and $T$ is the temperature in Kelvin.

The Lorentz half-width for HITRAN data is given as \footnote{https://hitran.org/docs/definitions-and-units/}
\begin{equation}
\label{eq_gammaL}
\Gamma_L = \frac{A}{4\pi c} + \left( \frac{T}{T_{ref}}\right)^{-n} \left[ \frac{\alpha_{air} (P-P_{self})}{P_{ref}} + \frac{\alpha_{self} P_{self}}{P_{ref}}\right],
\end{equation}
which is designed for studies of the terrestrial atmosphere, so broadening is considered in two parts: air-broadening and self-broadening.
For ExoMol data, the corresponding definition is \citep{ty+2016}
\begin{equation}
\Gamma_L = \frac{A}{4\pi c} + \alpha_{ref}\left( \frac{T_{ref}}{T} \right)^n \left( \frac{P}{P_{ref}}\right)
\end{equation}
where $\alpha_{\text{ref}}$ gives the line broadening for the given broadening gas under consideration.
In both cases, we have included the first term with the Einstein A coefficient (which has physical units of s$^{-1}$), but it is generally negligible. 

For atomic opacities, the Lorentz half-width consists mainly of a combination of the the natural broadening coefficient and pressure broadening due to collisions with electrons (Stark broadening) and other atomic and molecular species (van der Waals broadening). Stark broadening depends directly on the abundance of free electrons and is, therefore, only important in hot stellar atmospheres. In this work, we only consider natural broadening for atomic opacities.

\subsubsection{Natural broadening}
The natural broadening of atomic transition lines is a consequence of the uncertainty principle $\Delta E \Delta t \ge \hbar / 2$ (\citealt[p.~156]{Kunze2009}; \citealt[p.~57]{Draine2011}) and applies to both the upper and lower transition states. The line profile of the natural broadening effect can be described by a Lorentzian line profile with a half-width at half-maximum, 
\begin{equation}
    \Gamma_L = \frac{\Gamma_{\rm nat}}{4\pi},
\end{equation}
where $\Gamma_{\rm nat}$ is the sum over all Einstein A coefficients of transitions to the upper or  lower energy levels \citep[p.~57]{Draine2011},
\begin{equation}
    \label{eq_GammaRad}
    \Gamma_{{\rm nat},lu} = \sum_{E_j < E_u} A_{uj} + \sum_{E_j < E_l} A_{lj}.
\end{equation}
For example,
\[
    \Gamma_{{\rm nat},31} = A_{31} + A_{32} 
\]
or
\[
    \Gamma_{{\rm nat},32} = A_{31} + A_{32} + A_{21}.
\]

For Kurucz data, we use
\begin{equation}
\Gamma_L = \Gamma_{\rm nat} = \frac{10^{\Gamma_R}}{4 \pi c}.
\end{equation}
The quantity $\Gamma_R$ is tabulated in the Kurucz database.


\subsubsection{Line intensity}
Besides the Voigt line profile, the intensity $S$ of the transition line is also needed. It is defined in units of cm g$^{-1}$  as (e.g. \citealt{hyt13} or Chapter 5.3 of \citealt{Heng2017})
\footnote{Note that the equation in \cite{hyt13} is defined in units of cm/molecule, without the mass term in the denominator.}
\begin{equation}
\label{eq:S}
    S = \frac{A}{8 \pi c m} \frac{g'}{\tilde\nu_0^2 Q(T)} e^{\left(-\frac{hc}{k_B T}E''  \right)} \left( 1 - e^{\left(-\frac{hc}{k_B T}\tilde\nu_0  \right)} \right), 
\end{equation}
where $E''$ is the lower-state energy in $\text{cm}^{-1}$, $g'$ is the upper-state statistical weight, and $c=2.99792458 \times 10^{10}$ cm s$^{-1}$ is the speed of light.  Note that there is a typographical error in the first exponent of equation 17 of \cite{gh15}, which we correct here.

The partition function $Q(T)$ is in general available in a tabulated form, together with the line list data. If the partition function is not available, it may be computed using (e.g. \citealt{sb2007})
\begin{equation}
\label{eqn:partition}
Q(T) = \sum_{i = 1}^n g_i e^{-hcE_i /k_BT}, 
\end{equation}
where $g_i$ and $E_i$ are the statistical weight and the energy of the level $i$, respectively. 
Note that, as \cite{sb2007} wrote, this summation diverges formally for high pressure values. For the pressure values that we are typically interested in, the sum is still accurate enough.
For molecular species, the partition function would also include electronic, vibrational, and rotational terms \citep{sb2007}.

\subsubsection{Decompose the $a$-$u$ Voigt parameter plane}

\begin{figure}[h]
\begin{center}
\includegraphics[width=\columnwidth]{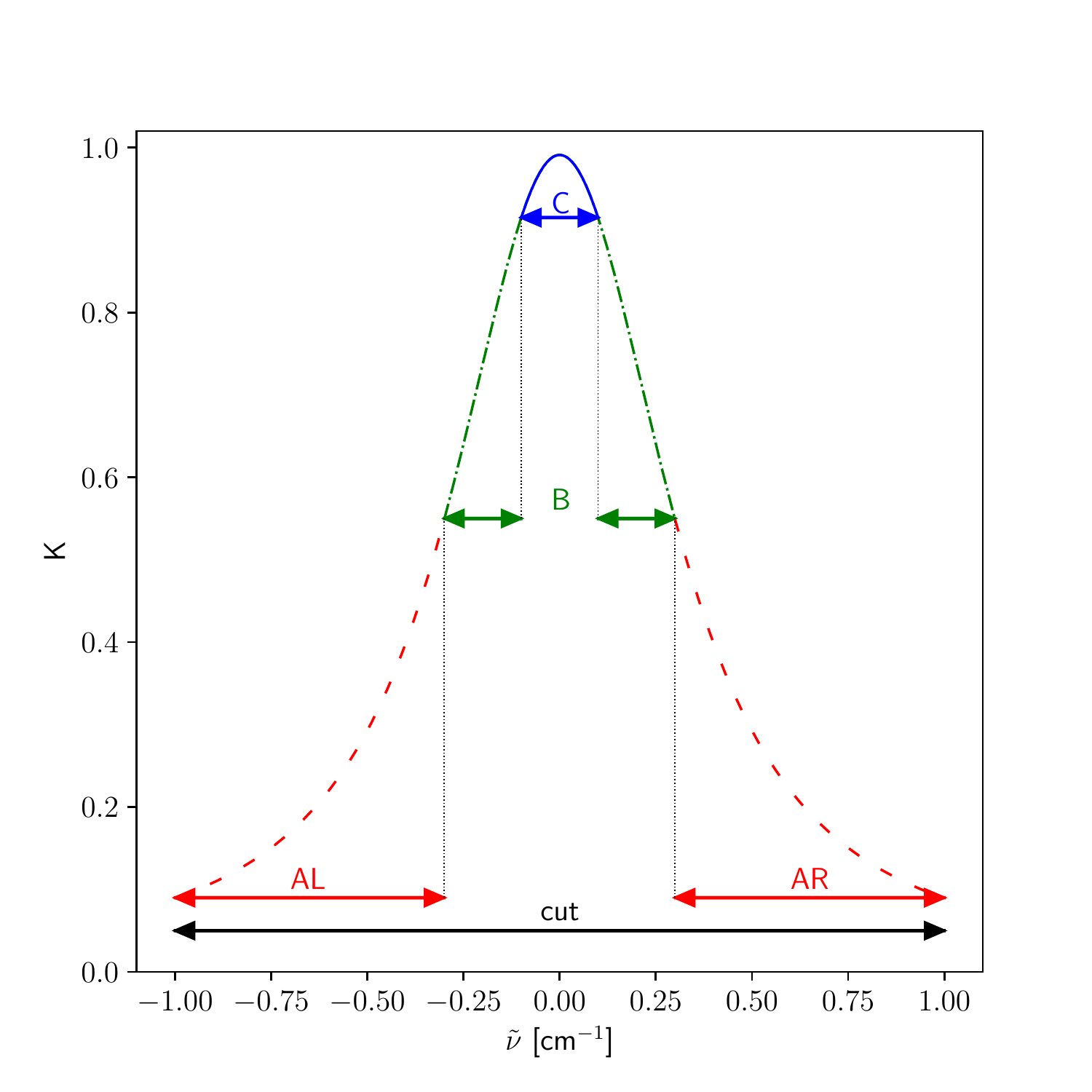}

\end{center}
\caption{Example of the $a$-$u$ regions, applied on a single transition line, with a cutting length of 1 cm$^{-1}$. In practice, thousands of neighboring transition lines are grouped together, and the regions, especially B and C, can overlap.}
\label{fig:regions}
\end{figure}

\begin{figure}[h]
\begin{center}
\includegraphics[width=\columnwidth]{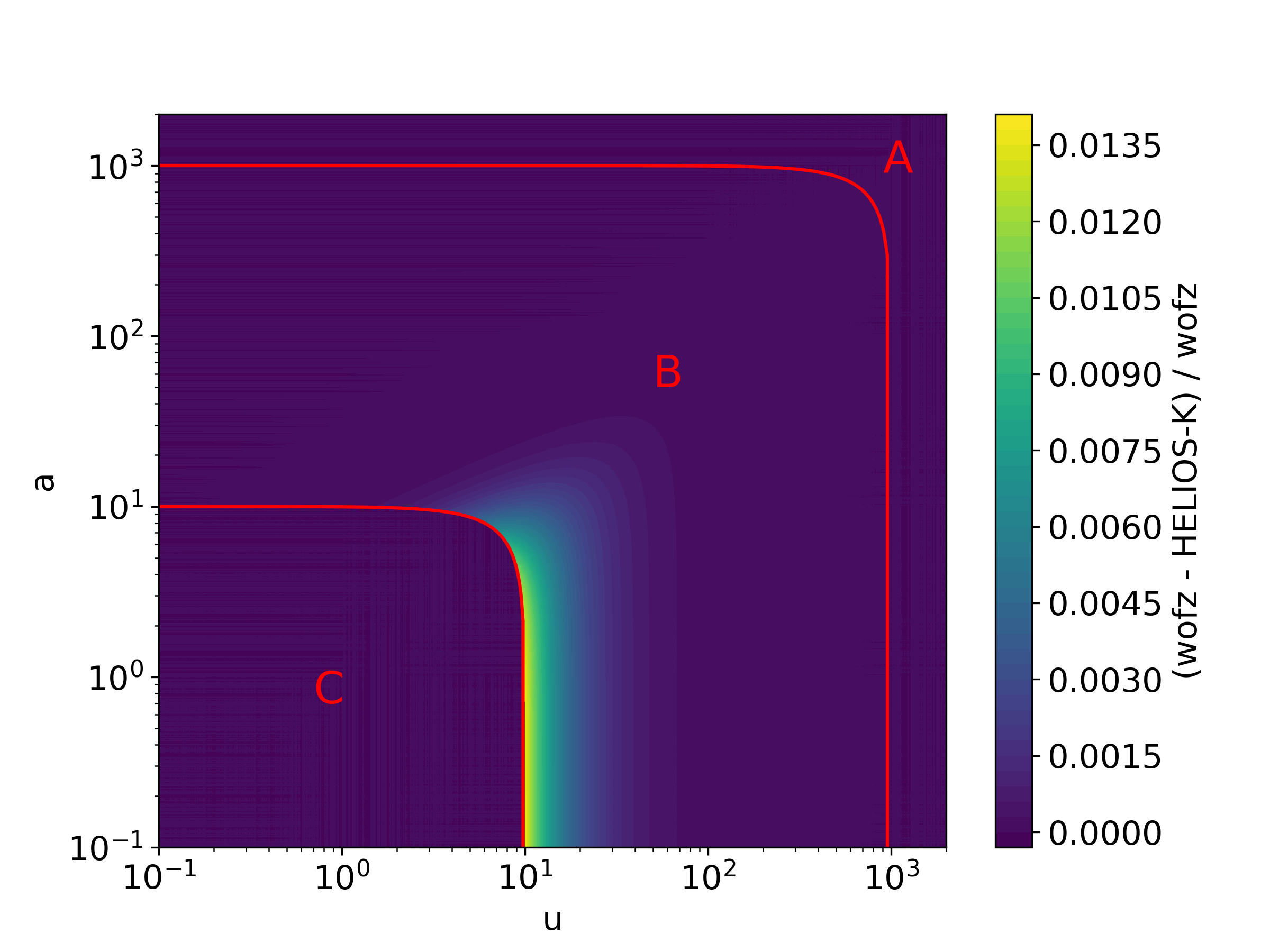}

\end{center}
\caption{Fractional difference of the Voigt profile computed by \texttt{HELIOS-K} and using the \texttt{SciPy} function \texttt{wofz} (\texttt{scipy.special.wofz}). The boundaries between regions A, B, and C from Table \ref{tab:regions} are shown in red. The algorithm in region C is similar to the \texttt{wofz} function, while regions B and A are approximations to speed up the calculation.}
\label{fig:regionsError}
\end{figure}

Following \cite{lb07} and as already described in \cite{gh15}, we split the $a$ and $u$ parameter space into three different regions, called A, B, and C, and use different methods and approximations for the Voigt profile calculations. For the A region, we use a first-order Gauss-Hermite quadrature approximation. For the B region, we use a third-order approximation.  For the C region, we solve the Voigt profile up to machine precision with an iterative method \citep{za11}. In practice, we split the A region into a left (AL) and a right (AR) part in order to avoid many zero calculations at the line center. We do not split the B region in the same way, because most of the time the C region is very narrow. Splitting B would introduce too many unnecessary memory accesses, especially when the $a$-$u$ regions are applied to thousands of neighboring transition lines, and the B and C regions can overlap. The used $a$-$u$ regions are listed in Table \ref{tab:regions} and illustrated in Figure \ref{fig:regions}.
Figure \ref{fig:regionsError} shows the fractional difference between the \texttt{HELIOS-K} Voigt calculation and the \texttt{SciPy} function \texttt{wofz} (\texttt{scipy.special.wofz}). The \texttt{wofz} function is similar to the algorithm in our region C. The largest error appears at the inner boundary of region B, where the Voigt function is approximated by a third-order Gauss-Hermite quadrature. The precision of \texttt{HELIOS-K} could be improved if necessary by expanding region C to larger values on the $u$-axis.

\begin{table}[h]
\centering
\begin{tabular}{c|c|c}

 $a$-$u$ Region & Limits & Method\\
 \hline
 AL & $ 10^6 < u^2 + a^2$       & First order Gauss- \\
   & $u < 0$  & Hermite quadrature  \\
 AR & $ 10^6 < u^2 + a^2$       & First order Gauss- \\
   &  $u > 0$ & Hermite quadrature  \\
 B & $ 100 < u^2 + a^2 < 10^6$ & Third order Gauss- \\
   & & Hermite quadrature \\
 C & $ u^2 + a^2 < 100$        & Algorithm 916 \\
   & & \citep{za11} \\

\end{tabular}
\caption{The $a$-$u$ plane is split into different regions, where different approximations on the Voigt profile calculations are applied. We find this splitting to be very efficient for GPU calculations, where branch divergencies should be avoided, and memory access should be minimized.}
\label{tab:regions}
\end{table}

Note that \cite{lb07} split the C region into different subregions with higher-order quadrature methods or interpolation of precalculated tables. But we find that by using GPUs, the three regions from table \ref{tab:regions} provide a good balance between the efficiency of the method, parallel applicability, and resource limitations on the GPU.

\subsection{GPU implementation considerations}
\label{sect:GPU}
Using GPUs for numerical calculations is a very powerful approach. Modern GPUs can consist of more than 5000 CUDA cores, which allow us to parallelize the numerical problem at a high level. However, there are some limitations. First, not all CUDA threads are independent, and threads in the same warp (a collection of 32 threads) must perform the same operation but can use individual data (single instruction, multiple threads, SIMT). This means that thread divergences must be avoided. In our problem, thread divergences could occur when neighboring transition lines lie in different $a$-$u$ regions and require a different algorithm to calculate the Voigt profile. To avoid those divergences, we have to sort the data and organize them in a well-suited form. 

A second limitation of the GPUs is the memory transfer bottleneck. Before data can be used on the GPU, it must be transferred to the GPU through the peripheral component interconnect (PCI) express, which can cause large computing overhead time. However, the data transfer time can be hidden by using asynchronous and interleaved memory transfer. This is possible because the GPUs have separated copy and compute engines \footnote{Nvidia GeForce cards typically use the same copy engine for data transfer for host-to-device and for device-to-host. Tesla cards have separate copy engines for host-to-device and for device-to-host and can work simultaneously.}, and both can work simultaneously. Besides the memory transfer, CUDA also allows us to overlap kernel execution with calculations performed on the CPU. In our problem, we can therefore hide data reading and data transfer behind the opacity calculations on the GPU. But that only works if the GPU part takes long enough. That is typically the case when using high wavenumber resolution calculations ($\Delta \tilde\nu \approx 0.01$ cm$^{-1}$). If this is not the case, then the full computing time is dominated by pure data reading.

A further limitation is the slow global memory access. Nvidia GPUs have different levels of memory, mainly, global memory, shared memory, and registers. Most of the time, memory access goes through global memory first and is then stored temporarily in shared memory or registers. Global memory access is slow and should be avoided. This can be achieved by designing the parallelization scheme in such a way that data can be reused multiple times from registers or shared memory.

\subsection{Improvements and Upgrades to the \texttt{HELIOS-K} opacity calculator}
\label{improvments}

This section describes details of the parallelization and new implementations of \texttt{HELIOS-K}. Readers who are not interested in these coding details can skip this section and continue with the different line list databases in section \ref{sect:prepare}.

\subsubsection{Reading the line list data}

Before the opacity calculation can be started, all of the line list data need to be read from the CPU and transferred to the GPU. We use a binary data format to store preprocessed line lists to speed up this reading process. How the line lists are preprocessed is described later in section \ref{sect:prepare}.
To read the data from the hard disk sounds simple, but actually, one has to be very careful not to lose too much performance. In fact, when using low-to-moderate resolution in wavenumber, reading the line list can take more time than the opacity calculation itself. When the code is called many times, e.g. for different values of temperature and pressure, reading and writing of data can become the bottleneck. We use the following strategy to improve the reading and data transfer process.
\begin{itemize}
	\item Avoid reading data line by line. Reading the line list, e.g. by an \textit{fscanf} command line by line, would not generate enough work to hide the overhead time, and the entire memory bandwidth would not be used. Instead, we read a bigger block of data, consisting of a few thousand transition lines, at once by using the \textit{fread} command and a memory buffer. 
	\item Use asynchronous and interleaved memory copy to transfer the data to the GPU. In this way, one block of data is transferred to the GPU at the same time as the CPU reads in the next block. 
	\item Overlap the reading and memory transfer with the opacity calculation of the previous block of data. Here we use the fact that the compute engine can work simultaneously with the copy engine and that the CPU can work in parallel to the GPU (see section \ref{sect:GPU}). In order to make this work, a clever scheduler of the opacity kernel calls and memory transfer commands is needed. This is described in detail in section \ref{sect:scheduler} and illustrated in Figure \ref{fig:stream}.
\end{itemize}

\subsubsection{Code parameters for the Voigt profile calculation}
\label{sect:parameters}

In practice, we do not use the parameters in equations (\ref{eq_a}-\ref{eq_gammaL}) in their stated form but instead cancel out the $\sqrt{\ln 2}$ factor in the numerator and denominator and directly store the inverse of the Doppler half-width, because the calculation of a product is faster than a division. The parameters used are then
\begin{equation}
    x = (\tilde\nu - \tilde\nu_0) \cdot \alpha_D^{-1},
\end{equation}
and
\begin{equation}
    y = \Gamma_L \cdot \alpha_D^{-1},
\end{equation}
with 
\begin{equation}
\label{eq_alphaD}
\alpha_D^{-1} = \frac{c}{\tilde\nu_0} \sqrt{\frac{m}{2  k_B T}}.
\end{equation}
The parameters $x$, $y$, and $\alpha_D^{-1}$ correspond to the parameters $u$, $a$ and $\Gamma_D$ from equations (\ref{eq_a}) -- (\ref{eq_gammaD}). We introduce here a new notation for the parameters that closely follows the implementation in the code. Since an ideal code implementation reduces the number of operations in each equation to the minimum, it is sometimes necessary to redefine parameters to be slightly different from their theoretical derivations. 

When a constant resolution $\Delta\tilde{\nu}$ is used, the parameter $x$ can be calculated very fast with a single fused-multiply-add operation as
\begin{equation}
\label{eqn:x}
x = d + i \cdot b,
\end{equation}
with a thread index $i$, and the variables $d = (\tilde\nu_{\text{min}} - \tilde\nu_0) \cdot \alpha_D^{-1}$ and $b =\Delta \tilde\nu \cdot \alpha_D^{-1}$. The quantity $\tilde\nu_{\text{min}}$ is the starting point in wavenumber.

Assuming a constant resolution in wavenumber is a natural choice, because the wavenumber points nearly follow the positions of the transition lines. In \texttt{HELIOS-K}, it is possible to use a different wavenumber resolution for different bands. In that case, the variable $x$ in equation (\ref{eqn:x}) is stored in a precalculated array. 

In summary, the calculation of the Voigt profile needs a total of five parameters, which depend on the transition line and the wavenumber: the line intensity \textit{S}; the quantities \textit{b}, \textit{d}, and \textit{y} as defined above; and an additional quantity $\tilde{\nu}_{\text{cut}}$ that handles the cutting length. The result of the opacity calculation is written into an array \textit{K}. The major difficulty is how to handle these five parameters efficiently within GPU memory and use them in a parallel way.

\subsubsection{Calculating the line parameters}
As described in section \ref{sect:parameters}, we need five line parameters, \textit{S, b, d, y,} and $\tilde{\nu}_{\text{cut}}$. The calculation of these parameters is split into four different parts.
\begin{enumerate}
    \item The line intensity (Equation \ref{eq:S}) may be written as
    \begin{equation}
\label{eq:S1}
    S = \frac{S'}{Q(T)} e^{\left(-\frac{hc}{k_B T}E''  \right)} \left( 1 - e^{\left(-\frac{hc}{k_B T}\tilde\nu_0  \right)} \right),
\end{equation}
where the term $S'$ does not depend on the temperature or pressure,
\begin{equation}
\label{S1}
S' = \frac{g' A}{8 \pi c \tilde{\nu}^2 m}.
\end{equation}

    This quantity $S'$ can be calculated beforehand and is part of the preprocessing routine, which is described below in section \ref{sect:prepare}. In this step, the partition function $Q(T)$ has to be determined by reading and interpolating the tabulated values provided by the line list databases. 
    \item The next step takes place after a block of line list data is transferred into a memory buffer on the GPU. The parameters \textit{S and y} and, temporarily the inverse Doppler half-width $\alpha_D^{-1}$ are then calculated on the GPU. We use one thread per transition line for these calculations, and the quantities are stored in arrays in global memory on the GPU. 
    \item The line list data obtained are mostly sorted at this point but can be slightly out of order due to pressure shift effects. For that reason, we sort the quantities along $\tilde{\nu}$ using the CUDA \texttt{Thrust} library\footnote{https://developer.nvidia.com/thrust}. 
    \item In the last step, the quantities \textit{b, d} and
    $\tilde{\nu}_{\text{cut}}$ are calculated, where $\tilde{\nu}_{\text{cut}}$ depends on how the cutting length is defined. Options include a fixed cutting length, a fraction of the Doppler half-width, or a fraction of the Lorentzian half-width. 
    
\end{enumerate}
With this way of splitting the calculation of the parameters, we can reduce the sorting step to a minimum.

\subsubsection{Parallelizing the opacity calculations}
\label{sect:tiles}
The opacity calculation can be parallelized in two dimensions: the wavenumber $\tilde\nu$ and the transition line index $id$.

Parallelizing both dimensions simultaneously would not be efficient in terms of global memory data access. It is better to load a part from the data into registers and read it multiple times from there. Therefore, we divide the $\tilde\nu$-\textit{id} plane into data tiles and parallelize in the first step only the line index dimension. For the wavenumber dimension, we use a serial loop to access all points. In a second step, the serial loop can be split further into multiple threads.
How many threads to use for this process is dependent not only on the parameters but also on the GPU type and is therefore hard to estimate beforehand. For that reason, we use a self-tuning routine, which periodically checks for the best number of threads to use and adapts the CUDA-kernel calls automatically. 

\begin{figure}[h]
\begin{center}
\includegraphics[width=\columnwidth]{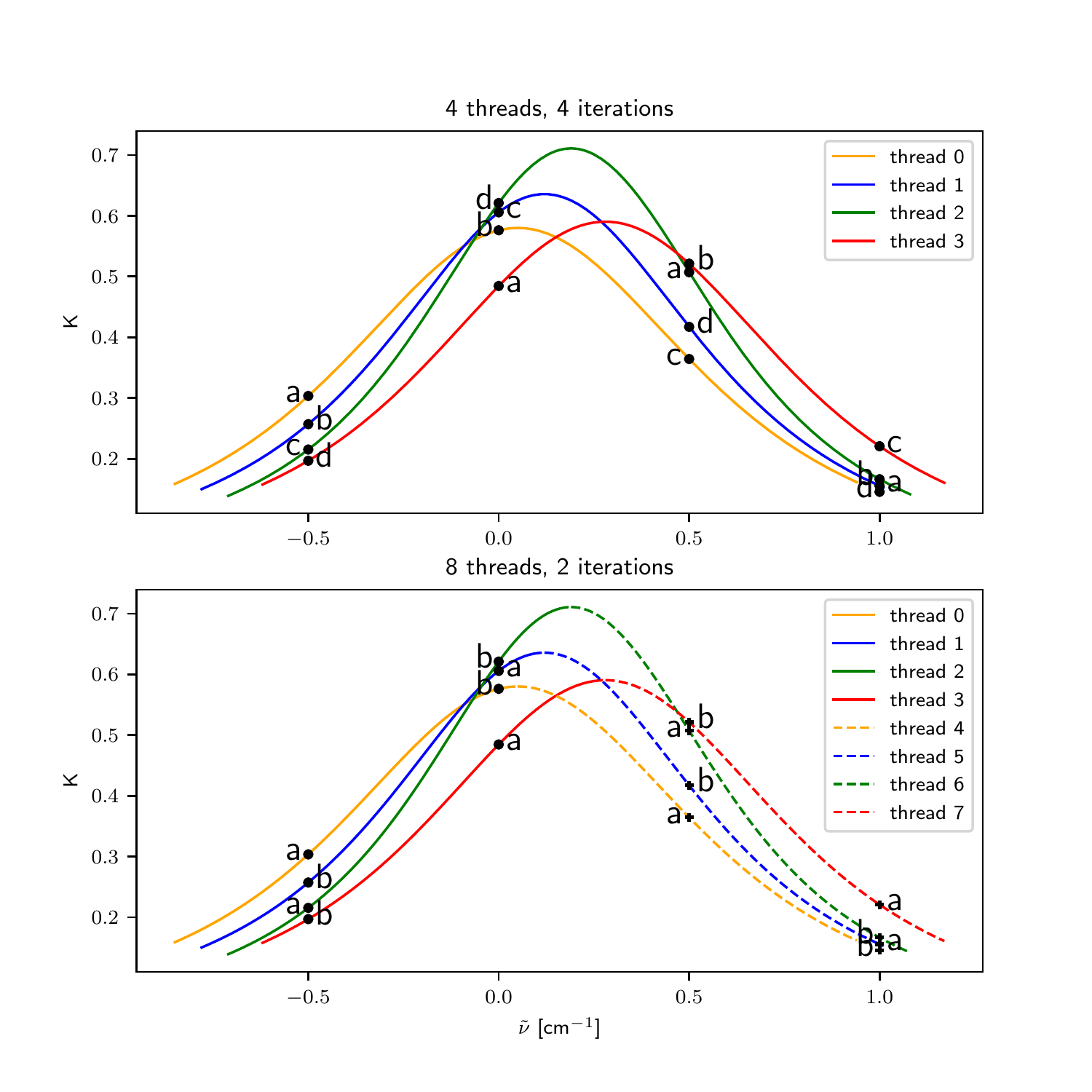}

\end{center}
\caption{Schematic for parallelizing the opacity calculation for an example with four transition lines and four points in wavenumber. In the top panel, four threads are used, and each thread has to iterate through points a, b, c, and d. In the bottom panel, eight threads are used, and each thread has to iterate through points a and b. In the second case, the results have to be collected in two different arrays to avoid race conditions. The transition lines are cut at a distance from the line center of 0.9 cm$^{-1}$.}
\label{fig:K}
\end{figure}

The top panel of Figure \ref{fig:K} gives a schematic of the parallelization scheme for an example with four transition lines and four points in wavenumber (-0.5, 0.0, 0.5, and 1.0). When we use four threads in this example, each thread is assigned to a different transition line (different color in the Figure) and has to iterate four times around the points in wavenumber. It is important that each thread calculates a different wavenumber point during the same iteration, because otherwise it would generate a race condition, which would lead to incorrect results. To prevent that, we can shift the starting point by the line index and take the modulo operator as

\begin{align}
\tilde{\nu} \text{\_index = (line\_index + iteration\_-index})  \nonumber \\
\text{\% (number of threads).}
\end{align}

In the example from Figure \ref{fig:K}, e.g. the thread 2 would then calculate the opacities in the order (0.5, 1.0, -0.5, and 0.0).
It is also possible to use more than four threads for the example above, as shown for eight threads in the bottom panel of Figure \ref{fig:K}. But then we have to use separate arrays to collect the results of the different threads and merge the results later because, e.g. iteration b of thread 1 happens at the same time as iteration b of thread 3, and both threads write to the same point in wavenumber. Using eight threads instead of four increases the amount of parallel work performed by the GPU but also increases the fraction of reading vs. computing, which means global memory access cannot be hidden behind the calculation time. Which version is the fastest depends on the individual case. Our algorithm self-tunes to find the best option.  

Listing \ref{lst:kernel} shows a pseudocode for a GPU kernel that reads the transition line parameters and calls the Voigt profile function. The parameter \texttt{NBx} sets the number of iterations each thread has to perform. The parameter \texttt{nl} is the number of transition lines in a block of data. The parameters \texttt{il1} and \texttt{nstart} define the starting points of the transition line index and wavenumber index of the current block of data. The kernel is launched on a two-dimensional grid. The x-dimension of that grid is defined as the maximum range of wavenumber points of the current data block divided by \texttt{NBx}. The y-dimension is used to launch multiple blocks of data together in order to reduce the kernel launch overhead time. The y-dimension is on the order of 16 or 32. 

In lines 12 - 16 of Listing \ref{lst:kernel}, the five transition line parameters from section \ref{sect:parameters} are loaded from global memory into registers. The values in those registers can be reused \texttt{NBx} times. The result of the opacity calculation is stored in a two-dimensional shared memory array, \texttt{K\_s}. At the end of the calculation, in lines 30 - 35, the results from multiple threads are merged into a single array, which can then be written back into global memory and at the very end copied back to the CPU.

\begin{lstlisting}[caption={Pseudocode of the opacity calculation kernel.},captionpos=b,label=lst:kernel,basicstyle=\small,frame = single]

Line_kernel(){
  int idx = threadIdx.x;
  int idy = blockIdx.x * NBx;

  int il = il1 + blockIdx.y * nl;

  for(int iil=0;iil < nl;iil += blockDim.x){

    int iL = iil + il + idx;
        
    S = S_d[iL];
    y = y_d[iL];
    d = d_d[iL];
    b = b_d[iL];
    cut2 = cut2_d[iL];
                
    for(int i = idx; i < NBx; ++i){
      ii = i % NBx;
      iii = nstart + idy + ii;
      x = d + iii * b;
      if(cut2 < x*x + y*y){
        K_s[i % (NBy * NBx)] += S*voigt(x,y);
      }
      __syncthreads();
    }
  }
  __syncthreads();

  if(idx < NBx){
    for(int j=1;j < blockDim.x / NBx;++j){
      K_s[idx] += K_s[idx + j*NBx];
      __syncthreads();
    }
  }
}
\end{lstlisting}

\subsubsection{Calculating the $\tilde\nu$ limits}

As described before, the transition lines are grouped together into blocks of data, where each block consists of a few thousand lines. In order to call the Voigt calculation kernel, we have to determine the range of wavenumbers of each block. This range is defined as either the maximum or the minimum of the cutting range or the limits of the $a$-$u$ regions described in Table \ref{tab:regions}. The limits $\Delta \tilde \nu$ can then be calculated with the equation
\[
a^2 + u^2 = l^2,
\]
where $l$ is either 10 or 1000, for the regions B and C. Inserting equations (\ref{eq_a}), (\ref{eq_u}), and (\ref{eq_alphaD}) leads to
\[
 \frac{\Gamma_L^2}{\alpha_D^2} + \frac{(\tilde\nu - \tilde\nu_0)^2}{\alpha_D^2} = l^2
\]
and
\begin{equation}
\Delta \tilde\nu = (\tilde\nu - \tilde\nu_0) = \sqrt{l^2 \cdot \alpha_D^2 - \Gamma_L^2}.
\end{equation}
Therefore, the limits of the B and C regions for an individual transition line are $\tilde\nu_0 \pm \Delta \tilde\nu$. The limit of the A region is simply $\tilde\nu_0 \pm$ $\tilde{\nu}_{\text{cut}}$, where $\tilde{\nu}_{\text{cut}}$ can depend on the Doppler or Lorentz half-width. 
In order to know the $\tilde\nu$ limits of an entire block of transition lines, we use a parallel reduction method on the GPU by using warp shuffle instructions.

\subsubsection{The core workflow}
\label{sect:scheduler}
\begin{figure*}[ht]
\begin{center}
\includegraphics[width=2.0\columnwidth]{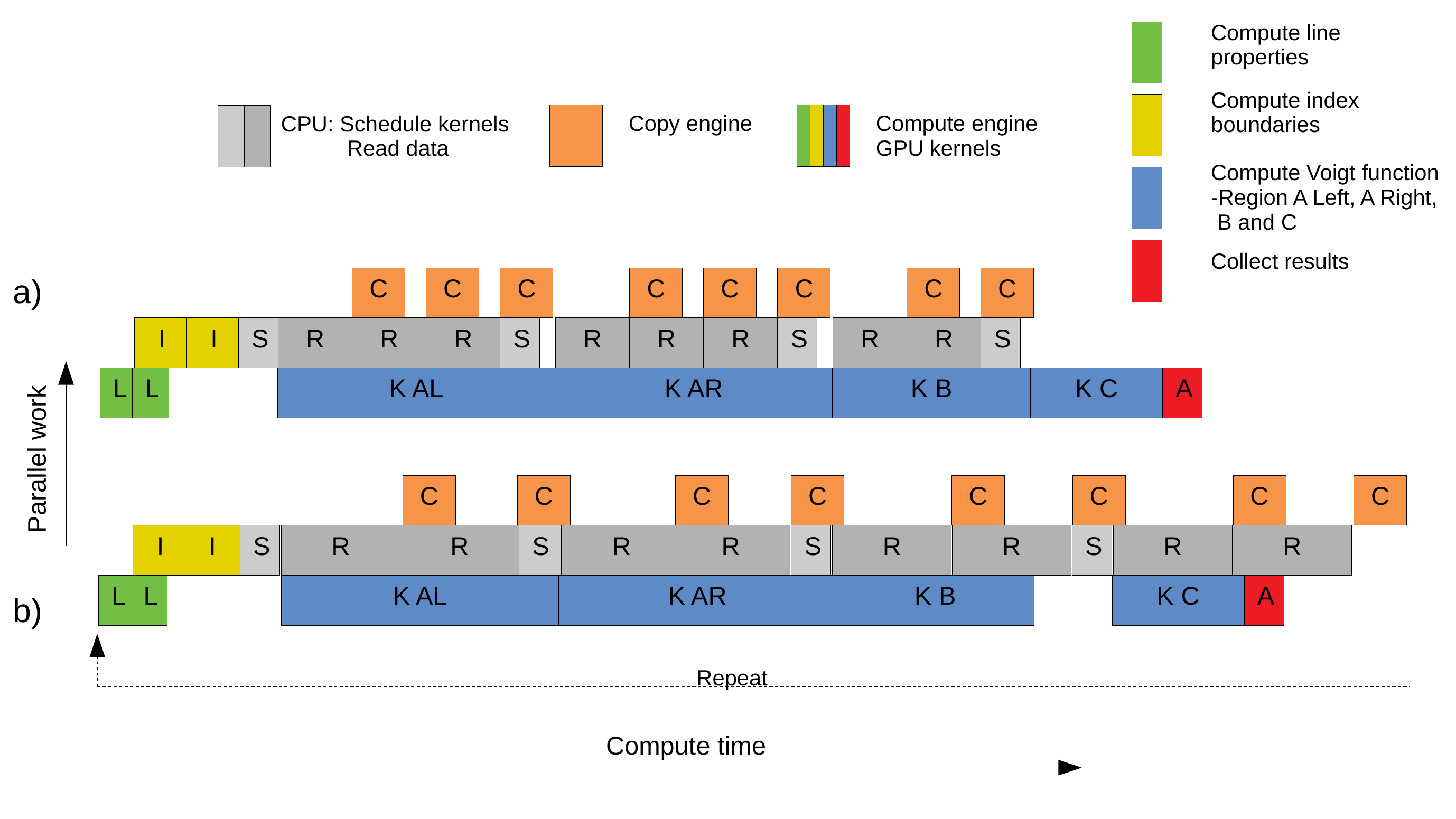}
\end{center}
\caption{Simplified workflow of the code for two cases: (a) fast memory access and (b) slow memory access. The boxes indicate the compute time for all involved operations on the CPU (light and dark gray) and the GPU (colors). When boxes are stacked on top of each other, the two operations are executed in parallel at the same time. In case (a) the compute time is dominated by the opacity calculation and in case (b) it is dominated by reading the data from the hard disk. Case (b) should be avoided.}
\label{fig:stream}
\end{figure*}

Typically, the calculation of the Voigt function is the longest part of the GPU calculation, and we want to use that execution time to simultaneously read the next block of data from the hard disk and transfer it to the GPU. This is possible because different GPU engines can work at the same time and GPU kernel calls are asynchronous, meaning that the workflow goes back to the CPU immediately after the kernel is called and does not wait for completion. (See section \ref{sect:GPU}.) The resulting workflow is shown in figure \ref{fig:stream} and can be described as the follows.
\begin{itemize}
    \item First, the line properties have to be calculated. This operation scales with the number of transition lines and is split into several different kernels, but it can be described mainly with two operations: 1) calculate the line intensity (``L"; green box in Figure \ref{fig:stream}) and (2) organization of the tiles described in section \ref{sect:tiles} (``I"; yellow boxes). The ``I" operation has to wait until ``L" is finished, but by using multiple streams, parts from ``L" and ``I" can overlap on the GPU. This is indicated in Figure \ref{fig:stream} by stacking boxes on top of each other.
    \item The calculation of the Voigt profile (blue boxes) is split into four different parts (A left, A right, B, and C; Table \ref{tab:regions}). Before the calculation starts, the CPU determines the total number of threads for the next block of data by using the information from the ``I" operation. This part is indicated in the figure as ``S." Immediately after ``S," the GPU kernel starts, and the CPU can be used for the next part. At the end of the four ``K" parts, temporary results are collected (``A"; red boxes). 
    \item While the GPU is busy, we use the CPU to read future data blocks from the hard disk. We have to interrupt this operation as soon as the ``K" kernel on the GPU is finished. In practice, we use a CUDA event query for this check.
    \item The new data are transferred asynchronously to the GPU (``C"; orange boxes), and the execution time is hidden almost completely behind the reading operation or GPU execution time.
    \item During the kernel execution, the self-tuning routine is called periodically to optimize the performance of the GPU, not shown in Figure \ref{fig:stream}.
\end{itemize}

The overall performance highly depends on whether the data reading is faster (case (a)) or slower (case (b)) than the Voigt profile calculation in the GPU (Figure \ref{fig:stream}). Case (b) can happen when the resolution in wavenumber $\Delta \tilde\nu$ is small, the GPU is very fast, or the data access from the hard disk is limited. Data access speed can generally be improved by moving the data closer to the calculation unit and ideally keeping data files in memory. If this is not possible, then opacities for multiple points in pressure should be calculated at once in order to reduce memory access. Some performance results using different data storage locations are listed in section \ref{performance}.

\subsection{Preparation of spectroscopic line list data for computation}
\label{sect:prepare}

For the opacity calculations, we use line lists from several databases, which all use a different data format and support a different way of downloading the data. In order to use \texttt{HELIOS-K} in the most efficient way, we define a uniform binary data format for the line lists and provide scripts for automatically downloading the data and converting them to this format. Besides the data format, we also define a parameter file for each species, which contains the metadata of the species. 

Our procedure is to process all transition lines from the given line lists. We do not use a truncation of weak lines. The line intensities are calculated in double precision floating point numbers. However, the final opacity function is reduced to single precision to reduce storage space, which truncates the opacity function to about $10^{-50} - 10^{-40}$ cm$^2$ g$^{-1}$. If a truncation of weak lines is required to reduce the total number of transition lines, it could easily be included in the \texttt{HELIOS-K} preparation scripts.

In this section, we describe the file structures that we use for \texttt{HELIOS-K} and outline how the data from different databases can be preprocessed and used. 

\subsubsection{Parameter files for species}
For each species (atom or molecule), we define a parameter file, which contains all necessary metadata and file structure information. These parameter files can be produced automatically and allow \texttt{HELIOS-K} to process the given data in a simple way, especially when the data from a database is split into many different subfiles. An example of a species-parameter file for HITRAN \ch{H2O} is given in Table \ref{tab:SParam}. The file lists all isotopologues or isotopes of the species with the corresponding abundances, molecular masses, and partition function file names. 

\begin{table*}[t]
\begin{lstlisting}[frame = single]
Database = 0
Molecule number = 1
Name = hit16
Number of isotopologues = 7
#ID Abundance       Q(296K)   g     Molar Mass(g)  Partition File :
11  0.997317        174.58    1     18.010565      q1.txt
12  0.002000        176.05    1     20.014811      q2.txt
13  3.718840E-04    1052.14   6     19.01478       q3.txt
14  3.106930E-04    864.74    6     19.01674       q4.txt
15  6.230030E-07    875.57    6     21.020985      q5.txt
16  1.158530E-07    5226.79   36    20.020956      q6.txt
17  2.419700E-07    1027.80   1     20.022915      q129.txt
Number of columns in partition file = 2
Number of line/transition files = 1
Number of lines per file :
304225
Line file limits :
0
30000
#ExoMol :
Number of states = 0
Number of columns in transition files = 0
Default value of Lorentzian half-width for all lines = 0
Default value of temperature exponent for all lines = 0
Version = 0

\end{lstlisting}
\caption{Example Species-parameter file for HITRAN \ch{H2O}. The \texttt{HELIOS-K} species-parameter files contain all relevant meta parameters necessary for the opacity calculation. The files can be generated automatically with the provided scripts.}
\label{tab:SParam}
\end{table*}

\subsubsection{Binary File Structure}
The different databases provide their data files in different structures and also with a different number of parameters. We define a binary data format for the atomic or molecular transition lines. Depending on the database, these binary files contain different values, as indicated in Table \ref{tab:binary1}. The possible values are described in Table \ref{tab:binary2}. In the binary files, we preprocess the data in such a way that the final opacity calculation can be performed as fast as possible. One important part is the precalculation of the temperature- and pressure-independent part of the line intensity $S'$, defined in Equation (\ref{S1}).

\begin{table}
\centering
\begin{tabular}{ |c|c|}
 \hline
 Database & Parameters \\
 \hline
 HITRAN, HITEMP & $ID$, $\tilde\nu$, $S'$, $E_L$, $A$, $\delta$, $\gamma_{Air}$, $\gamma_{Self}$, $n$ \\ 
 ExoMol         & $\tilde\nu$, $S'$, $E_L$, $A$ \\
 Kurucz, VALD3         & $\tilde\nu$, $S'$, $E_L$, $\Gamma_{\text{nat}}$ \\
 NIST           & $\tilde\nu$, $S'$, $E_L$, $A$ \\
 \hline
\end{tabular}
\caption{Values contained in the \texttt{HELIOS-K} binary data files, depending on the database used. The parameters are described in Table \ref{tab:binary2}.}
\label{tab:binary1}
\end{table}

\begin{table}
\centering
\begin{tabular}{ |c|c|c|c|} 
 \hline
 Value & Description & Unit \footnote{More details can be found at https://hitran.org/docs/definitions-and-units/} & Type \\
 \hline
 ID & Species identity &  & char4 \\ 
 $\tilde\nu$ & Center of the line & cm$^{-1}$ &  F64 \\
 $S'$ &  Equation \ref{S1} & cm g$^{-1}$ &  F64 \\
 $E_L$ & Lower energy level & cm$^{-1}$ & F64 \\
 $A$ & Einstein Acoefficient & s$^{-1}$ & F64 \\
 $\delta_{Air}$ & Pressure-dependent line shift & cm$^{-1}$ atm$^{-1}$ & F64 \\
 $\gamma_{Air}$ & Air broadening coefficient & cm$^{-1}$ atm$^{-1}$ & F64 \\
 $\gamma_{Self}$ & Self broadening coefficient & cm$^{-1}$ atm$^{-1}$ & F64 \\
 $n$ & Exponent of the temperature & & F64 \\
  & -dependent Lorentz half-width & & \\
 $\Gamma_\text{nat}$ & Natural broadening coefficient & s$^{-1}$ & F64 \\
 \hline
\end{tabular}
\caption{Possible values in the \texttt{HELIOS-K} binary data files.}
\label{tab:binary2}
\end{table}

\subsubsection{ExoMol}
The ExoMol line lists  \citep{Tennyson2020} can be accessed from \texttt{www.exomol.com} and are organized by molecules, isotopologues, and line list names. \texttt{HELIOS-K} provides a Python script to scan the webpage and produce a list of all available species. The species from ExoMol can be identified by their full species name, which includes the isotopologue information and the line list name, e.g. ``1H2-16O\_\_BT2". Each species has a ``.def" file, which contains properties like the isotopologue mass, default values of Lorentzian half-widths,  default values of temperature exponents, or the number of data files. It is recommended to check the number of files and maximal wavenumber of the line list. The data themselves are separated into a ``.states" file and one or multiple ``.trans" files. This data format is very compact to store on a server, but to use it with \texttt{HELIOS-K} we need to produce a new line list, which contains [$\tilde{\nu}$, $S'$, $E_L$, $A$]. 
This list is stored compactly as one or multiple binary files, which can be read by \texttt{HELIOS-K}. We provide a script that does the download and file conversion automatically. Some molecules contain the wavenumber of the molecular lines as a fourth column in the  ``.trans" files. However, it is not recommended to use that but instead to calculate it through the difference in energy levels. Partition functions are given in ``.pf" files.

For some species, ExoMol also provides precalculated cross-sections (xsec). When available, we compare our calculations with those data to verify our calculation (See section \ref{ExomlXsec}).

\subsubsection{ExoMol Superlines}
For some molecules, ExoMol provides superlines \citep{Yurchenko+2017,Tennyson2020}. Superlines are temperature-dependent collections of line intensities integrated over small wavenumber bins. By using superlines, the total number of Voigt profile calculations can be reduced dramatically. The drawback is that individual transition lines are not represented anymore, and all lines are calculated with the same default pressure-broadening values. Practically, it means that the quantities $S$ and $S'$ from equations 
(\ref{eq:S1}) and (\ref{S1}) do not need to be calculated anymore but can be read in directly. Only the mass term needs to be included in the superline intensities. For example, the use of superlines reduces the ExoMol POKAZATEL water line list from  $5.7 \times 10^9$ transition lines to $8.3 \times 10^6$ superlines.

\subsubsection{HITRAN and HITEMP}
The HITRAN database\footnote{\texttt{www.hitran.org}} \citep{HITEMP, HITRAN2016} provides line lists in standard formatted text files (*.par files). These files can be downloaded for a single isotopologue or a mix of different molecules and isotopologues. In addition to the line list, HITRAN provides partition functions and molecular metadata in a webpage table. HITRAN supports a programming interface, HAPI \citep{HAPI}, that allows data access through a Python script. However, several tests showed that the data in HAPI were not always updated; therefore, we opt to use the content of the webpage table directly.
For usage with \texttt{HELIOS-K}, we extract only the data needed from the line lists and store them in more compact binary files that contain [ID, $\tilde{\nu}$, $S'$, $E_L$, $A$, $\delta$, $\gamma_{Air}$, $\gamma_{Self}$, n].

\subsubsection{Kurucz}
\label{sect:Kurucz}
The Kurucz database\footnote{\texttt{http://kurucz.harvard.edu}} \citep{Kurucz2018,Kurucz2017} provides line lists and partition functions for a large number of neutral and ionized atoms. In this work, we use the GFNEW lines in wavenumber (\texttt{gfallwn08oct17.dat}), which includes isotope fractions and hyperfine splittings for some species. \texttt{HELIOS-K} provides a Python script to download the data file and the available partition functions and to extract the relevant data for the opacity calculations. The Kurucz database includes natural broadening (radiation dampening), as well as Stark and van der Waals broadening coefficients. 
The Stark broadening coefficient in the Kurucz database is given at a temperature of 10,000 K in units of the electron number density. The van der Waals coefficients, on the other hand, are also stated for a temperature of 10,000 K but in units of the atomic hydrogen number density. Van der Waals broadening in the Kurucz database is based on calculations employing a modified version of Uns{\"o}ld's approximation \citep{unsold1938physik, Aller1963aass.book.....A}. The tabulated coefficients can be adapted to collisions with He and \ch{H2} by scaling them with a factor of 0.42 or 0.85, respectively.

In the current version, we only include natural broadening in the opacity calculation. The resulting atomic opacities are therefore suitable for environments where collisional broadening is negligible, such as atomic absorption lines originating from upper atmospheres.

\paragraph{Partition functions}
For most species, partition functions are available from the Kurucz database. If this is not the case, then we calculate the partition functions according to equation (\ref{eqn:partition}) and the energy levels from the NIST database (see section \ref{NIST}).

\paragraph{Isotope fraction correction}
During the data processing, we noted an issue in the isotope fractions with the file \texttt{gfallwn08oct17.dat}. When a transition line consists of multiple isotopes but only one isotope is split up into its hyperfine structure, then the file contains an incorrect isotope fraction of the remaining isotopes. We show an example of a potassium transition line in Table \ref{tab:hyperfine}. Potassium consists of three natural isotopes:
$^{39}$K (93.3$\%$), $^{40}$K (0.0117$\%$), and $^{41}$K (6.7$\%$). In the example, isotope $^{39}$K is split in its hyperfine structure, while isotopes $^{40}$K and $^{41}$K are combined into an entry with an isotope fraction of 1.0, which is not correct. The combined isotope fraction should instead be 6.7117$\%$. If this factor is not corrected, then the resulting opacity can be significantly overestimated. Our \texttt{HELIOS-K} data preparation script scans all transition lines for these types of missing factors and corrects them. 

\begin{figure}[ht]
\begin{center}
\includegraphics[width=1.0\columnwidth]{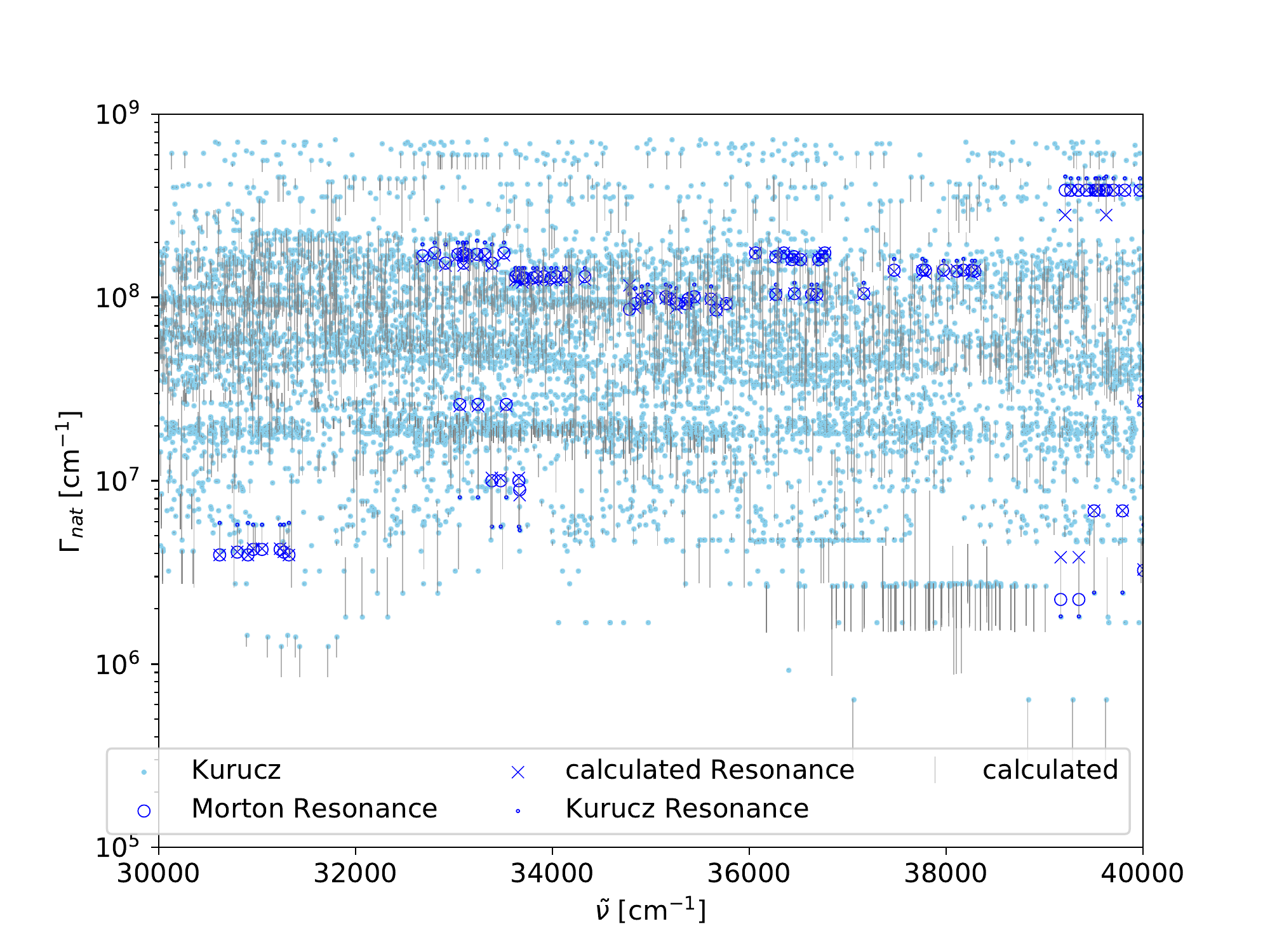}
\end{center}
\caption{Comparison of the natural broadening coefficients $\Gamma_{\text{nat}}$ of \ch{Fe} between tabulated values from Kurucz (light blue and dark blue dots), calculated values using the Einstein A coefficients (gray error bars and crosses), and published data \citep{Morton2003, Morton2008} for resonance lines (blue circles).}
\label{fig:Morton2}
\end{figure}

\paragraph{Natural broadening coefficients}
Natural broadening coefficients are provided for many species, but not for all. We compute the missing natural broadening coefficients according to 
equation (\ref{eq_GammaRad}). The natural broadening coefficient $\Gamma_{\text{nat},ij}$ between an upper state $i$ and a lower state $j$ includes the Einstein A coefficients from all transitions between two other levels, $k$ and $l$ with either

\begin{align}
\begin{cases}
    <E>_i = <E>_k \\
    g_i = g_k
\end{cases}
\end{align}
or
\begin{align}
\begin{cases}
    <E>_j = <E>_k \\
    g_j = g_k,
\end{cases}
\end{align}
where $<E>$ are the energy states averaged over the hyperfine sublevels and $g_{j,k}$ are the statistical weights from the Kurucz database. 
We compare our calculations to the tabulated values from Kurucz and with published values from resonance lines \citep{Morton2003, Morton2008}, as shown in Figure \ref{fig:Morton2} for \ch{Fe}. For most lines, our values correspond well with those from \cite{Morton2003, Morton2008}, but for a few transition lines, there are differences. Our calculated values are close to the ones from Kurucz, but for many lines, we cannot exactly reproduce them. The reason could be that we do not have full access to all relevant Einstein A coefficients or line configurations. The described comparison allows us to estimate the quality of calculated natural broadening coefficients for species where there are no tabulated values available.

\begin{table}[h]
\centering
\begin{tabular}{|c|c|c|c|c|}
\hline 
$\tilde{\nu}$ [cm$^{-1}$] & Isotope & Hyperfine Fraction & Isotope Fraction \\
\hline 
5554.11 & 39 & 0.031 & 0.933 \\
5554.11 & 39 & 0.437 & 0.933 \\
5554.11 & 39 & 0.062 & 0.933 \\
5554.11 & 39 & 0.156 & 0.933 \\
5554.11 & 39 & 0.156 & 0.933 \\
5554.11 & 0 & 1.0 & \textbf{1.0 (wrong)} \\
5554.11 & 39 & 0.156 & 0.933 \\
\hline 
\end{tabular}
\caption{Example of a Potassium transition line with a wrong isotope fraction in the Kurucz database. The missing fraction must be calculated to get correct opacities. The isotope fraction should be \textbf{0.067117}. Similar issues occur for several species and various transition lines. The numbers are rounded and taken from the file \texttt{gfallwn08oct17.dat} from \texttt{http://kurucz.harvard.edu}.}
\label{tab:hyperfine}
\end{table}

\subsubsection{NIST}
\label{NIST}
The NIST database\footnote{\texttt{https://www.nist.gov/pml/atomic-spectra-database}} \citep{NIST_ASD} provides line lists and energy levels for a large number of neutral and ionized atoms. The database can be accessed through a web interface by selecting the desired species and additional configuration options. The web access is intuitive and simple, but it is not very practical for accessing in an automated way, and it can be very time-consuming to download multiple line lists by hand.
Therefore, we provide a Python script that navigates the webpage automatically and fills in the necessary web forms to download the line lists and energy level files. The script also extracts the needed quantities from the files and converts them into the standard \texttt{HELIOS-K} format. 

A potential problem of the NIST database is that energy levels and transition lines can be listed multiple times. We noted this especially for the hydrogen atom, where energy levels are listed for individual angular momentum quantum numbers, besides the total averaged energy levels. Simply processing the entire line list would lead to duplicated transition lines; thus, it is necessary to filter the line lists for such duplicated entries. 

\paragraph{Partition functions}
We calculate the partition functions by the available energy levels according to equation (\ref{eqn:partition}). As \cite{sb2007} noted, this summation diverges formally for high pressure values. For pressure values that we are typically interested in, the sum is still accurate enough.

\paragraph{Natural broadening coefficients}
The natural broadening coefficients are not provided by the NIST database; therefore, we use equation (\ref{eq_GammaRad}) to calculate them by summing over all relevant Einstein A coefficients. The natural broadening coefficient $\Gamma_{\text{nat},ij}$ between an upper state $i$ and a lower state $j$ includes the Einstein A coefficients from all transitions between two other levels, $k$ and $l$ with either

\begin{align}
\begin{cases}
    \text{configuration}_i = \text{configuration}_k \\
    \text{term}_i = \text{term}_k \\
    J_i = J_k
\end{cases}
\end{align}
or
\begin{align}
\begin{cases}
    \text{configuration}_j = \text{configuration}_k \\
    \text{term}_j = \text{term}_k \\
    J_j = J_k,
\end{cases}
\end{align}
where ``configuration," ``term," and $J$ are given in the NIST database.

\begin{figure}[ht]
\begin{center}
\includegraphics[width=1.0\columnwidth]{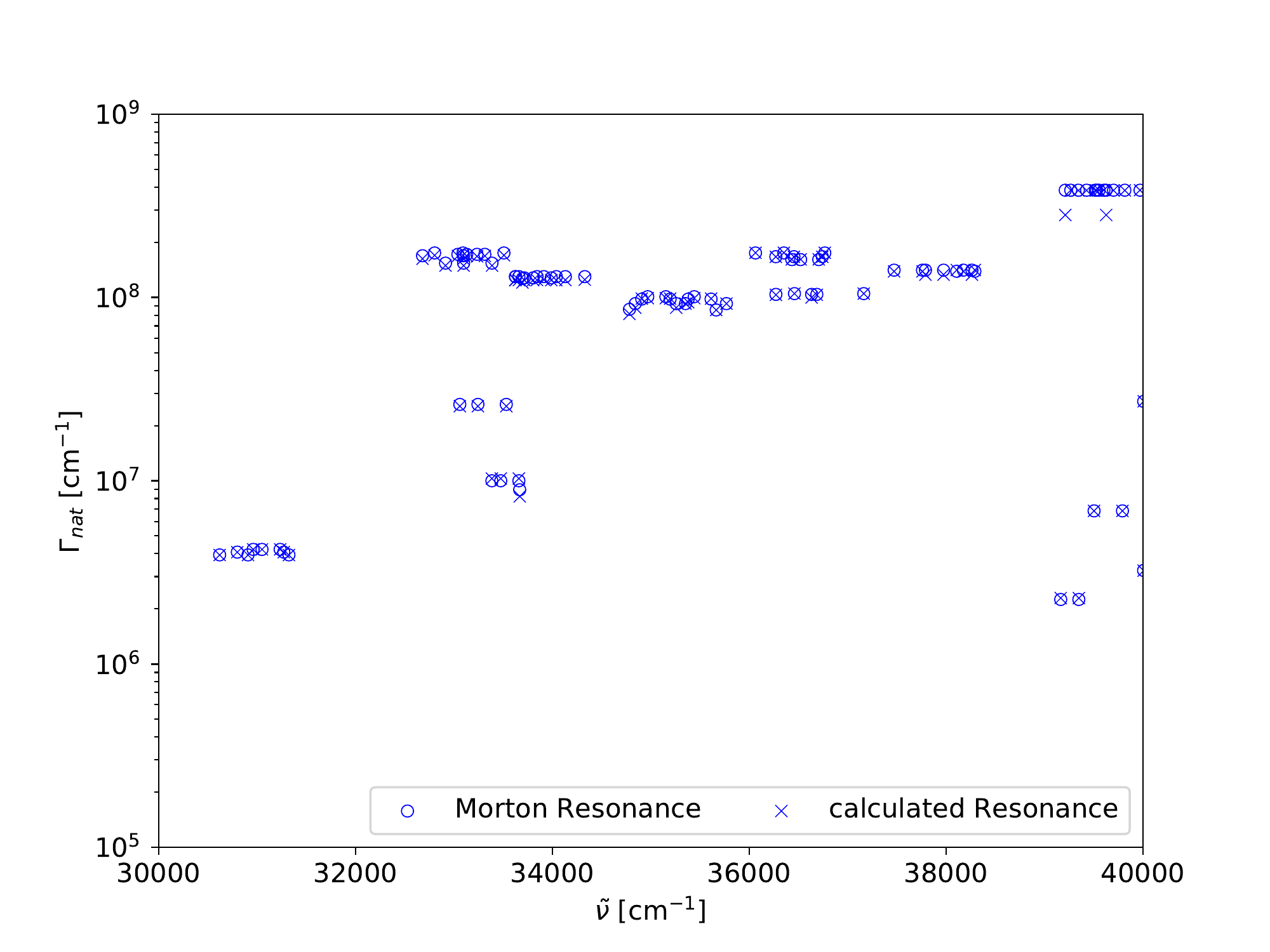}
\end{center}
\caption{Comparison of the natural broadening coefficients $\Gamma_{\text{nat}}$ of \ch{Fe} between calculated values from NIST data and published values from \cite{Morton2003}. Only resonance lines are shown. Our calculated values agree well for most transition lines.}
\label{fig:Morton}
\end{figure}

We compare our values with published values for resonance lines due to \cite{Morton2003, Morton2008} as shown in Figure \ref{fig:Morton} for \ch{Fe}. Differences between the NIST data and \cite{Morton2003, Morton2008} could occur because the underlying transition line data could be different. 

The preprocessing of the NIST data requires the following steps.
\begin{enumerate}
    \item Download the energy levels for all species. This can be done either manually through the web interface or by a script, which navigates the webpage automatically.
    \item Calculate the partition function. Alternatively, the partition function would be available through the web form but must be queried for each temperature individually. 
    \item Download the line lists for all species. Similar to the energy levels, this can be either done manually through the webpage or by automated script. 
    \item Calculate the natural broadening coefficients and generate the \texttt{HELIOS-K} binary files.
\end{enumerate}

\subsubsection{VALD3}

The VALD3 database\footnote{http://vald.astro.uu.se/} \citep{VALD2015, VALD2019} provides line lists for a large number of neutral and ionized atoms. After undergoing a registration process on the website, the database can be accessed through a web interface. Similar to our process for the NIST database, we provide a script to
navigate the website automatically. For every data request, VALD sends an email with a link to download the data. By requesting multiple species, this process can result in a large number of emails and data links. To simplify the download, our script can be used without opening all the sent emails, and it converts the line lists to the standard \texttt{HELIOS-K} format.

\paragraph{Partition functions}
Partition functions are not available in the VALD database. Therefore, we use the partition functions calculated from the NIST database.

\begin{figure}[ht]
\begin{center}
\includegraphics[width=1.0\columnwidth]{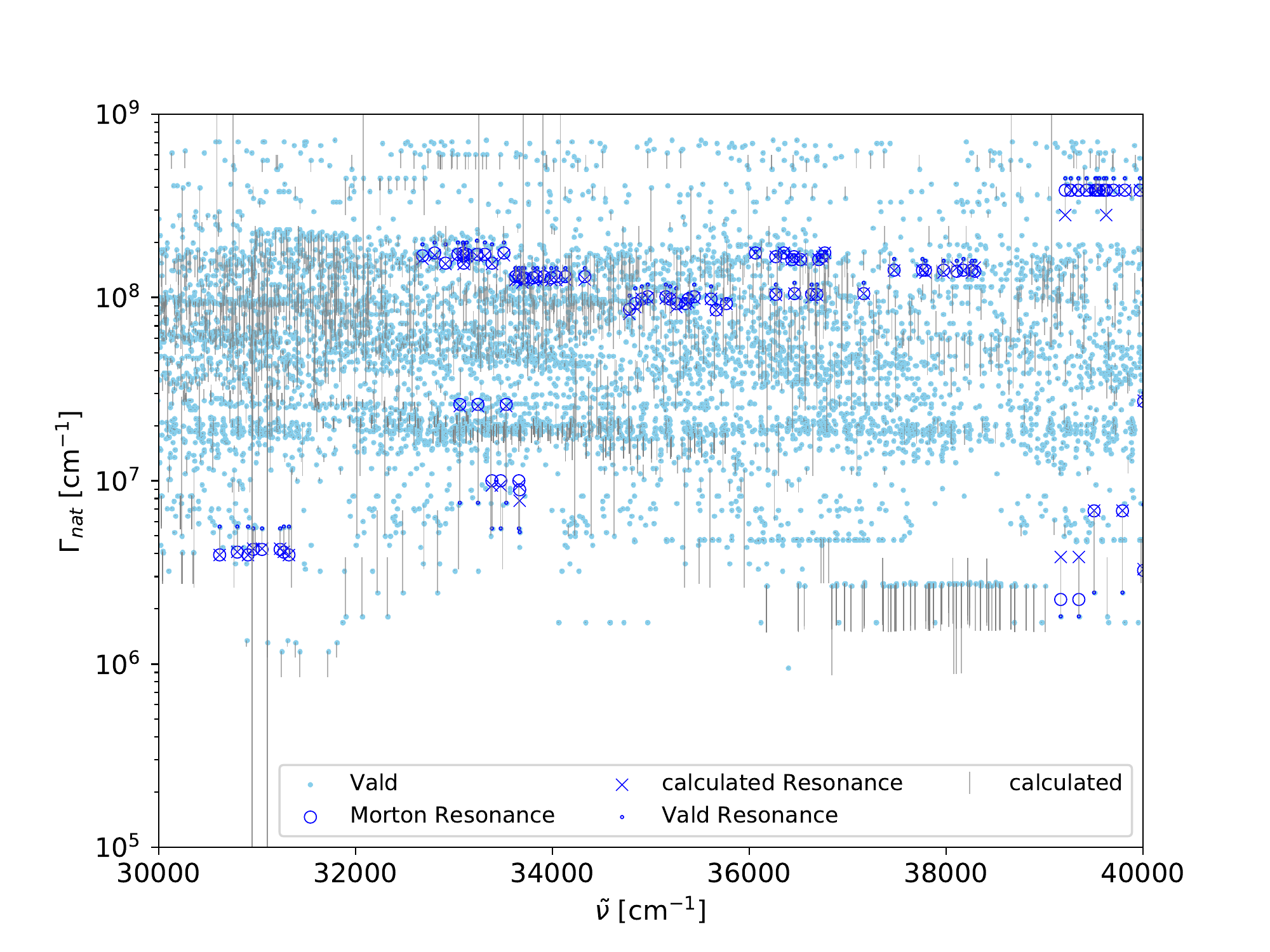}
\end{center}
\caption{Comparison of the natural broadening coefficients $\Gamma_{\text{nat}}$ of \ch{Fe} between tabulated values from VALD3 (light blue and dark blue dots), calculated values using the Einstein A coefficients (gray error bars and crosses), and published data \citep{Morton2003, Morton2008} for resonance lines (blue circles). In most cases, our calculated values correspond well with those from Morton, but for a few lines, there are differences.
}
\label{fig:Morton3}
\end{figure}

\paragraph{Natural broadening coefficients}
Natural broadening coefficients are provided for many species, but not for all. We compute the missing natural broadening coefficients as for the Kurucz database according to 
equation (\ref{eq_GammaRad}). The natural broadening coefficient $\Gamma_{\text{nat},ij}$ between an upper state $i$ and a lower state $j$ includes the Einstein A coefficients from all transitions between two other levels, $k$ and $l$ with either
\begin{align}
\begin{cases}
    E_i = E_k \\
    g_i = g_k
\end{cases}
\end{align}
or
\begin{align}
\begin{cases}
    E_j = E_k \\
    g_j = g_k,
\end{cases}
\end{align}
where the energy $E$ and the statistical weight $g$ are given by the VALD database. 
We compare our calculations to the tabulated values from VALD3 and with published values from resonance lines \citep{Morton2003, Morton2008} as shown in Figure \ref{fig:Morton3} for \ch{Fe}. For most lines, our values correspond well with those from Morton, but for a few transition lines, there are differences. Our calculated values are close to the ones from VALD3, but for many lines, we cannot exactly reproduce them. The reason could be that we do not have full access to all relevant Einstein A coefficients or line configurations. The comparison described allows us to estimate the quality of calculated natural broadening coefficients for species where there are no tabulated values available.

The preprocessing of the VALD3 data requires the following steps.
\begin{enumerate}
    \item Request the data download for all species. This can be done either manually through the web interface or by a script, which navigates the webpage automatically.
    \item Use the partition functions from the NIST database. 
    \item Calculate the natural broadening coefficients and generate the \texttt{HELIOS-K} binary files.
\end{enumerate}

\subsubsection{Alkali resonance lines}

The wings of the resonance lines of the alkali metals sodium and potassium are known to deviate from the usual Voigt profile. Especially their far-wing line profiles exhibit strong non-Lorentzian behavior due to collisions with other molecules, such as, in particular, H$_2$ and He. Various descriptions of these line profiles have been provided in the past to characterize these non-Lorentzian line wings \citep[see, e.g.][]{Tsuji1999ApJ...520L.119T, Burrows00, Burrows03, Allard2012A&A...543A.159A}. The most recent theoretical calculations of the resonance line wings broadened by collisions with H$_2$ were published by \citet{Allard2019A&A...628A.120A} for Na and by \citet{Allard2016A&A...589A..21A} for K. These calculations are valid for perturber densities of up to $10^{21}$ cm$^{-3}$.

\begin{figure}[ht]
\begin{center}
\includegraphics[width=1.0\columnwidth]{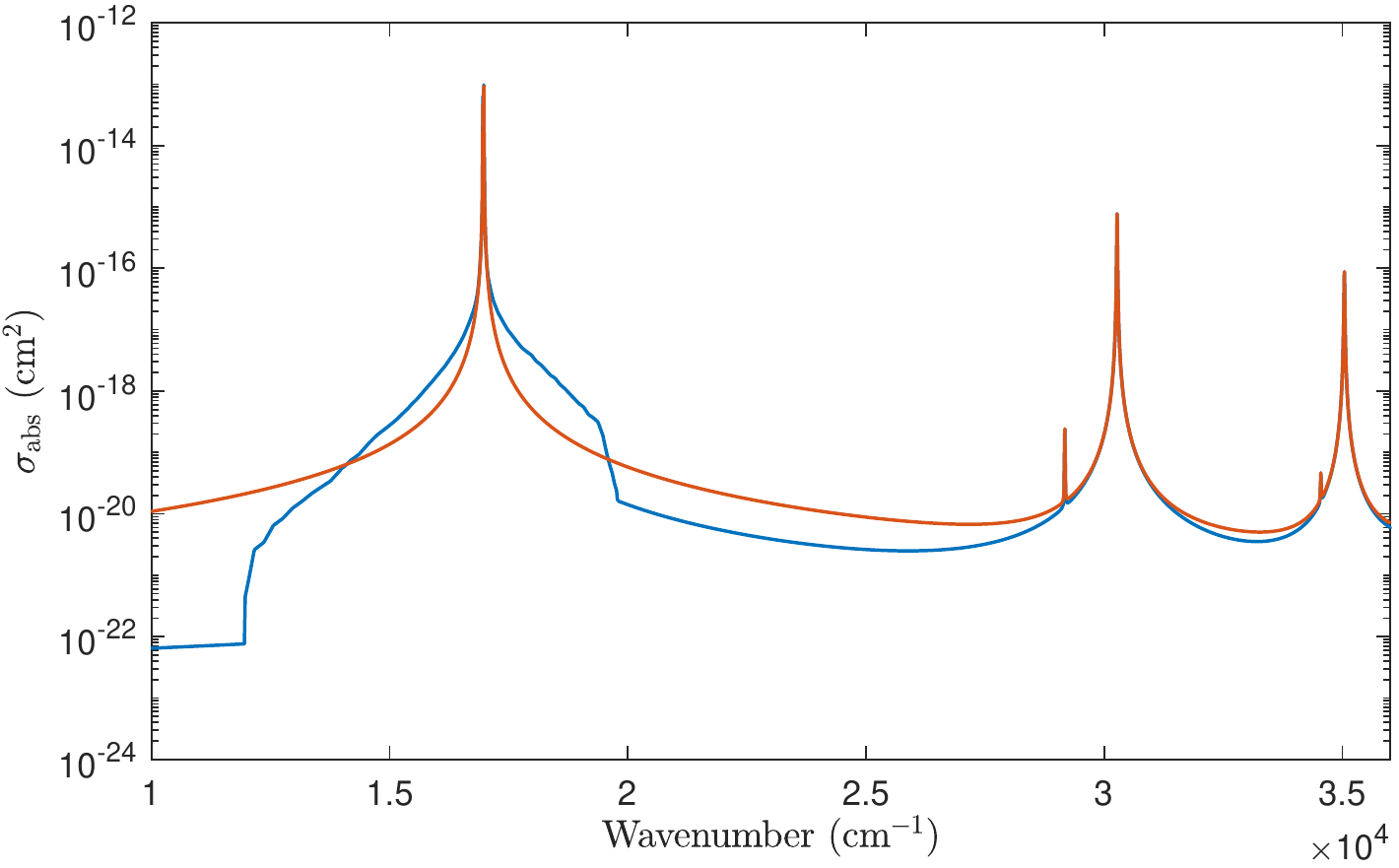}
\end{center}
\caption{Comparison of the absorption cross-coefficients of \ch{Na} between pure Voigt profiles (red line) and the non-Lorentzian line wings for the resonance lines based on \citet{Allard2019A&A...628A.120A} (blue line). The cross sections are calculated from the Kurucz line list data, including van der Waals broadening, and shown for a temperature of 800 K and an \ch{H2} partial pressure of 10 bars.
}
\label{fig:AllardSodium}
\end{figure}

Figure \ref{fig:AllardSodium} shows an example of the non-Lorentzian behavior of the sodium resonance line wings at a temperature of 800 K and an \ch{H2} partial pressure of 10 bars. We use the Kurucz line list to generate the cross sections, including the pressure-dependent van der Waals broadening by \ch{H2}. To convert the the corresponding broadening parameters listed in the Kurucz line list from H to \ch{H2}, we use a scaling factor of 0.85. For the calculation involving the \citet{Allard2019A&A...628A.120A} line wing description, we remove the two resonance lines from the line list, calculate the cross sections without them, and then add the two broadened resonance lines that have been constructed using the data from \citet{Allard2019A&A...628A.120A} to the results. 

The results shown in Figure \ref{fig:AllardSodium} clearly suggest that the resonance lines have a very strong non-Lorentzian behavior and are distinctively asymmetric. They are super-Lorentzian close to the line center and become sub-Lorentzian in the far wings. Using a Voigt line profile for these two lines thus results in overestimating the line absorption in the far wings by orders of magnitude, while strongly underestimating it near the line centers.

\section{Results}
\label{results}

\subsection{Comparison of atomic opacities}

We compare the atomic opacities between the databases NIST, Kurucz, and VALD3. As mentioned in section \ref{sect:Kurucz}, the calculations neglect pressure broadening. Therefore, the resulting opacities are suitable only for low-pressure environments with negligible collisional broadening. An example of four species (\ch{H}, \ch{Li}, \ch{S}, and \ch{Fe}) is shown in Figure \ref{fig:atomic}. 
Figures with more species are shown in appendix \ref{appendixA} (Figures \ref{fig:atomic0} - \ref{fig:atomic3}) and also in the \texttt{HELIOS-K} documentation.\footnote{\texttt{https://helios-k.readthedocs.io/en/latest/index.html}}
The 1H example of Figure \ref{fig:atomic} shows a good agreement between all three databases. The only difference is in the natural broadening coefficients from NIST. In the example of 3Li, the agreement is not as good as for 1H, and some of the smaller transition lines are different in all three databases. The more intensive transition lines agree better with the exception of the natural broadening coefficient of the largest line. The example of \ch{S} shows that NIST has far fewer lines than Kurucz and VALD3, and between Kurucz and VALD3, there is a large difference in the natural broadening coefficients around $\tilde{\nu}=84000$ cm$^{-1}$ . The overall agreement for \ch{Fe} is better, but again, NIST has far fewer lines, and Kurucz has more lines than VALD3.
In general, NIST has significantly fewer lines than Kurucz and VALD3. Kurucz and VALD3 are similar for many species, but still they are still not identical. For some species, VALD3 has fewer lines than Kurucz. The greatest impact on the opacities comes from the natural broadening coefficients from large resonance lines.

It should be noted that use of a  Voigt profile for atomic lines is not always suitable. Autoionization lines (Auger transitions) are a particular example that is not well described by a Voigt profile. The wings of such a line are more appropriately described by a so-called Fano profile \citep{Fano1961PhRv..124.1866F}, which decreases much faster than a Voigt profile. The Fano profile, however, depends on a parameter $q$ that cannot be easily calculated (see \citealt{Merts1972ApJ...177..137M} for details).
Therefore, using a Voigt profile for these lines will result in an overestimate of the absorption, creating an artificial continuum. This can be seen, for example, in Figure \ref{fig:atomic} for sulfur (atomic number 16). The opacities based on the Kurucz line list show an elevated continuum due to autoionization lines located in the far-UV wavelength region. Since these lines are not included in the VALD3 and NIST databases, their opacities are not affected by the overestimated line wings.

Other problematic lines are resonance lines, where using a normal Voigt profile can also lead to over- or underestimated absorption. In particular, the Lyman $\alpha$ wings of atomic hydrogen shown in the top left panel of Figure \ref{fig:atomic} are again overestimated by using a Voigt profile. Instead of following a Voigt profile, the actual line profile converges toward the Rayleigh scattering opacity far away from the line center \citep{Dijkstra2019SAAS...46....1D}.

\begin{figure}[ht]
\begin{center}
\includegraphics[width=1.0\columnwidth]{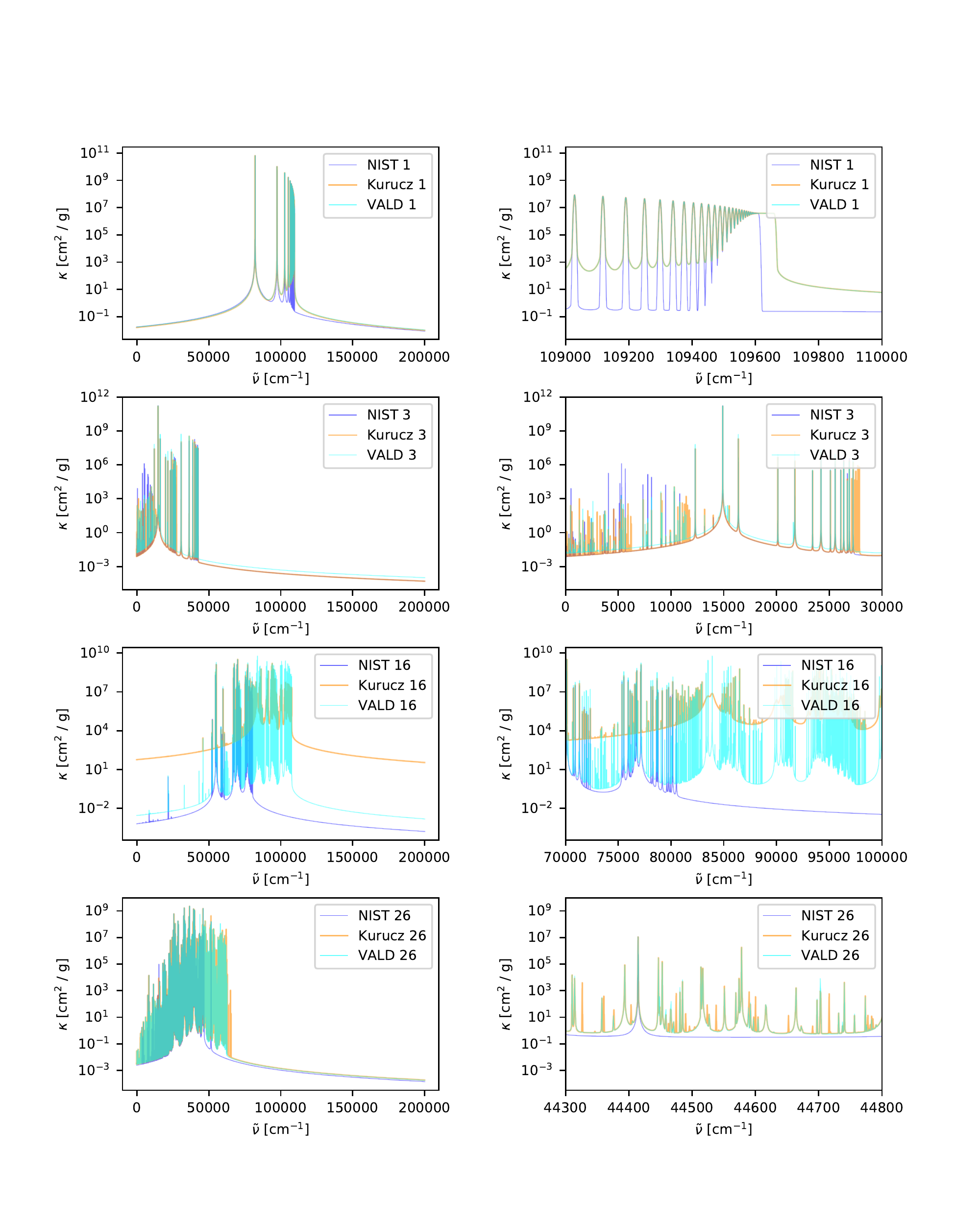}
\end{center}
\caption{Opacity comparison of four atoms (\ch{H}, \ch{Li}, \ch{S}, and \ch{Fe}) between the databases NIST, Kurucz, and VALD3. The panels in the left column show the entire wavenumber range, the panels in the right column show a more narrow band. We use a temperature of 3000 K and no cutting length. The three databases contain a different number of transition lines, leading to differences in wavenumber coverage of the opacities. Figures of all other species may be found in the \texttt{HELIOS-K} documentation.}
\label{fig:atomic}
\end{figure}

\subsection{Validation against ExoMol Opacities}
\label{ExomlXsec}
\begin{figure}[ht]
\begin{center}
\includegraphics[width=1.0\columnwidth]{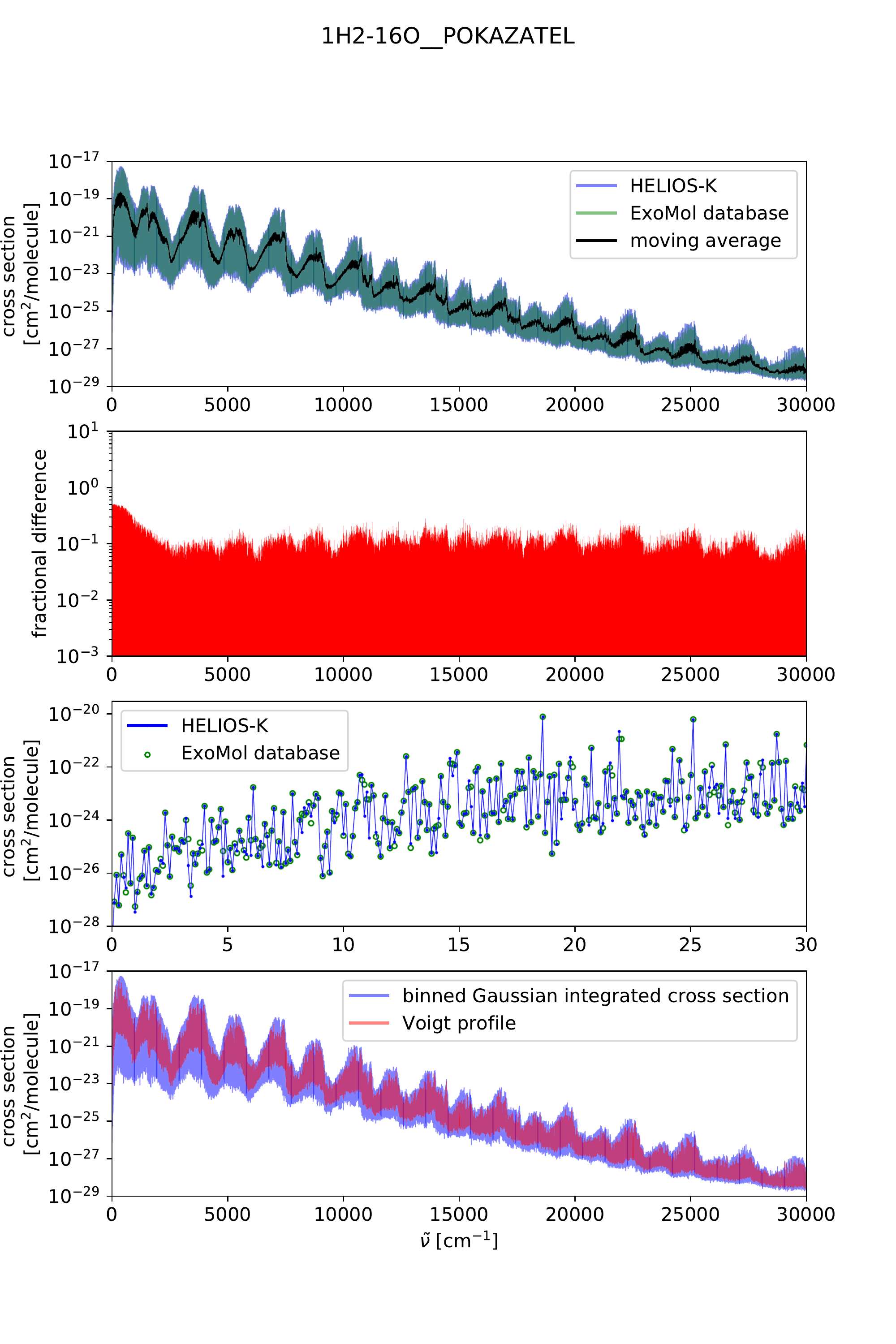}
\end{center}
\caption{First panel: comparison of our calculated cross sections of the POKAZATEL line list with cross sections provided from the ExoMol database, showing a good agreement.  Second panel: relative error between our work and ExoMol cross sections.  Third panel: zoom comparison between our work and cross sections from ExoMol. Individual points in wavenumber are not reproduced exactly due to different orders in summation processes.  Fourth panel: comparison between the binned Gaussian integrated cross section and the full Voigt profile.
Calculations are done with a resolution of 0.01 cm$^{-1}$ and a cutting length of 100 cm$^{-1}$. 
}
\label{fig:pokazatel}
\end{figure}

\begin{figure}[ht]
\begin{center}
\includegraphics[width=1.0\columnwidth]{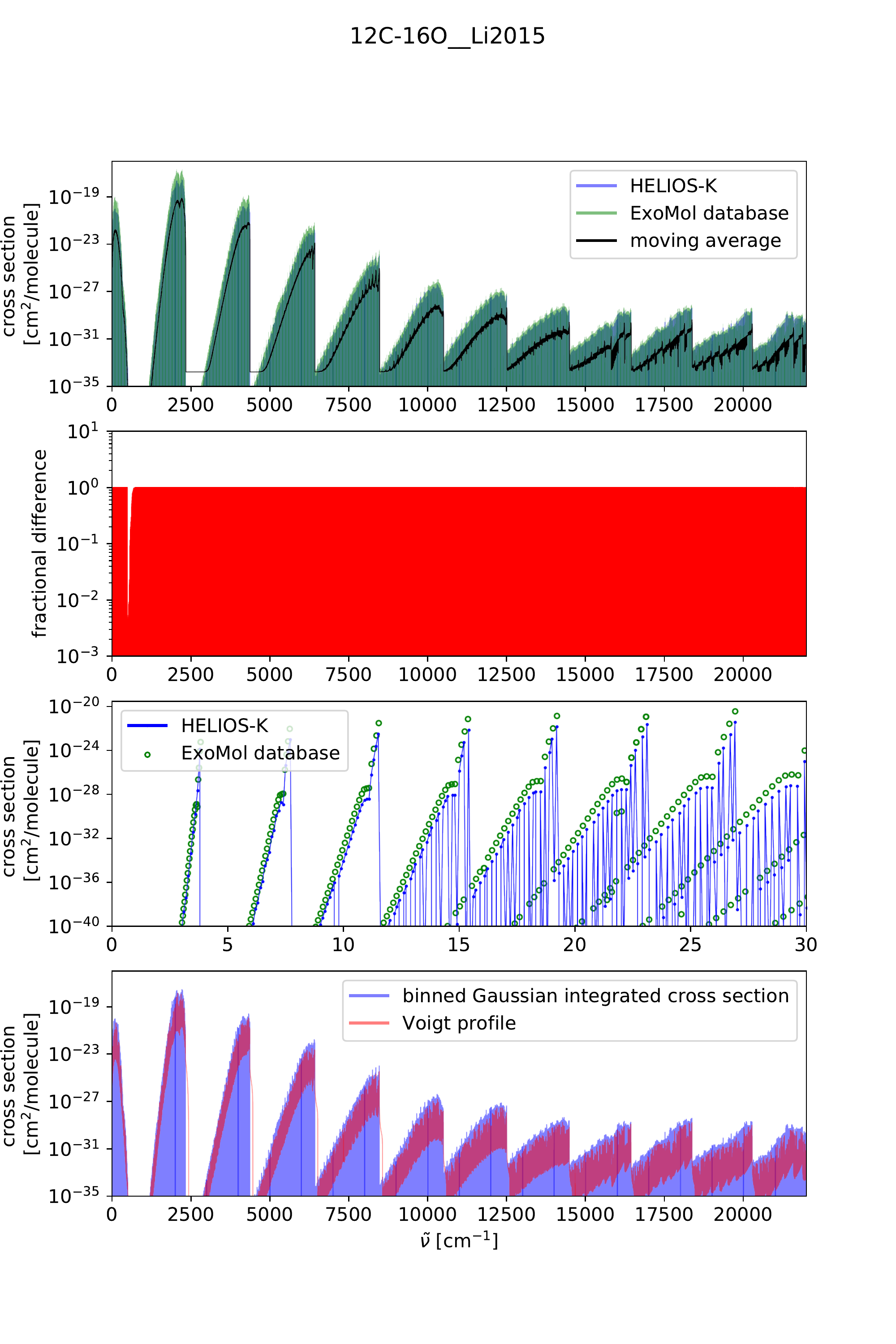}
\end{center}
\caption{Same as Figure \ref{fig:pokazatel} but for CO.
}
\label{fig:li2015}
\end{figure}

For many species, ExoMol provides precalculated binned Gaussian cross sections, which we use to validate our calculations. Even if the validation is not done on a full Voigt profile, it confirms that our line list preprocessing and line intensity calculations are done correctly. In Figure \ref{fig:pokazatel}, we show an example comparison of the \ch{H2O} POKAZATEL line list \citep{ExoMol2018}; and Figure \ref{fig:li2015}shows the same for CO \citep{ExoMol_CO}.
Individual points in wavenumber are not reproduced exactly, as shown in the bottom panel. One reason is that every point in wavenumber can consist of millions of individual transition lines, for which the intensity can vary by several orders of magnitude. If the order of summation over all the transitions involved is changed, it can affect the differences.

In appendix \ref{appendixB} (Figures \ref{fig:C0} - \ref{fig:C2}), we show a comparison for 45 molecules between binned Gaussian cross sections from ExoMol and computed with \texttt{HELIOS-K}. The fractional difference ($\kappa_{\text{HELIOS-K}} - \kappa_{\text{ExoMol}}) / \kappa_{\text{HELIOS-K}} $ typically reaches values between $10^{-7}$ and $10^{-1}$, which shows a good agreement between the two calculations. The difference can also reach higher values, which is most likely caused by division of very small numbers and limited numerical data precision. 

Examples of molecular opacities (full Voigt profile) for a temperature of 1500 K (500 K for SO$_3$), a pressure of 0.001 bar, and cutting length of 100 cm$^{-1}$ are shown in the Figures \ref{fig:fa} - \ref{fig:fd}. The figures include
TiO \citep{ExoMol_TiO}, VO \citep{ExoMol_VO}, AlO \citep{ExoMol_AlO}, SiO \citep{ExoMol_SiO},
CN \citep{ExoMol_CN}, CH \citep{ExoMol_CH}, CP \citep{ExoMol_CP}, CS \citep{ExoMol_CS},
H$_2$O \citep{ExoMol2018}, CO$_2$ \citep{HITEMP}, CO \citep{HITEMP, ExoMol_CO},
PN \citep{ExoMol_PN}, PO \citep{ExoMol_PO_PS}, PS \citep{ExoMol_PO_PS},
CH$_4$ \citep{ExoMol_CH4}, NH$_3$ \citep{ExoMol_NH3}, C$_2$H$_2$\citep{HITRAN2016}, HCN \citep{ExoMol_HCN_a, ExoMol_HCN_b},
SO$_2$ \citep{ExoMol_SO2}, SO$_3$ \citep{ExoMol_SO3}, SH \citep{ExoMol_NS_SH}, H$_2$S \citep{ExoMol_H2S},
AlH \citep{ExoMol_AlH}, SiH$_4$ \citep{ExoMol_SiH4}, SiH \citep{ExoMol_SiH}, SiS \citep{ExoMol_SiS}, 
NS \citep{ExoMol_NS_SH}, NH \citep{ExoMol_NH}, NO \citep{ExoMol_NO}, and NO$_2$ \citep{HITEMP, HITEMP2019}.

During our analysis, we also spotted several issues in the ExoMol cross sections or data files, which have since been corrected.

\subsection{Comparison of ExoMol Superline Opacities to Full Opacities}
\begin{figure}[ht]
\begin{center}
\includegraphics[width=1.0\columnwidth]{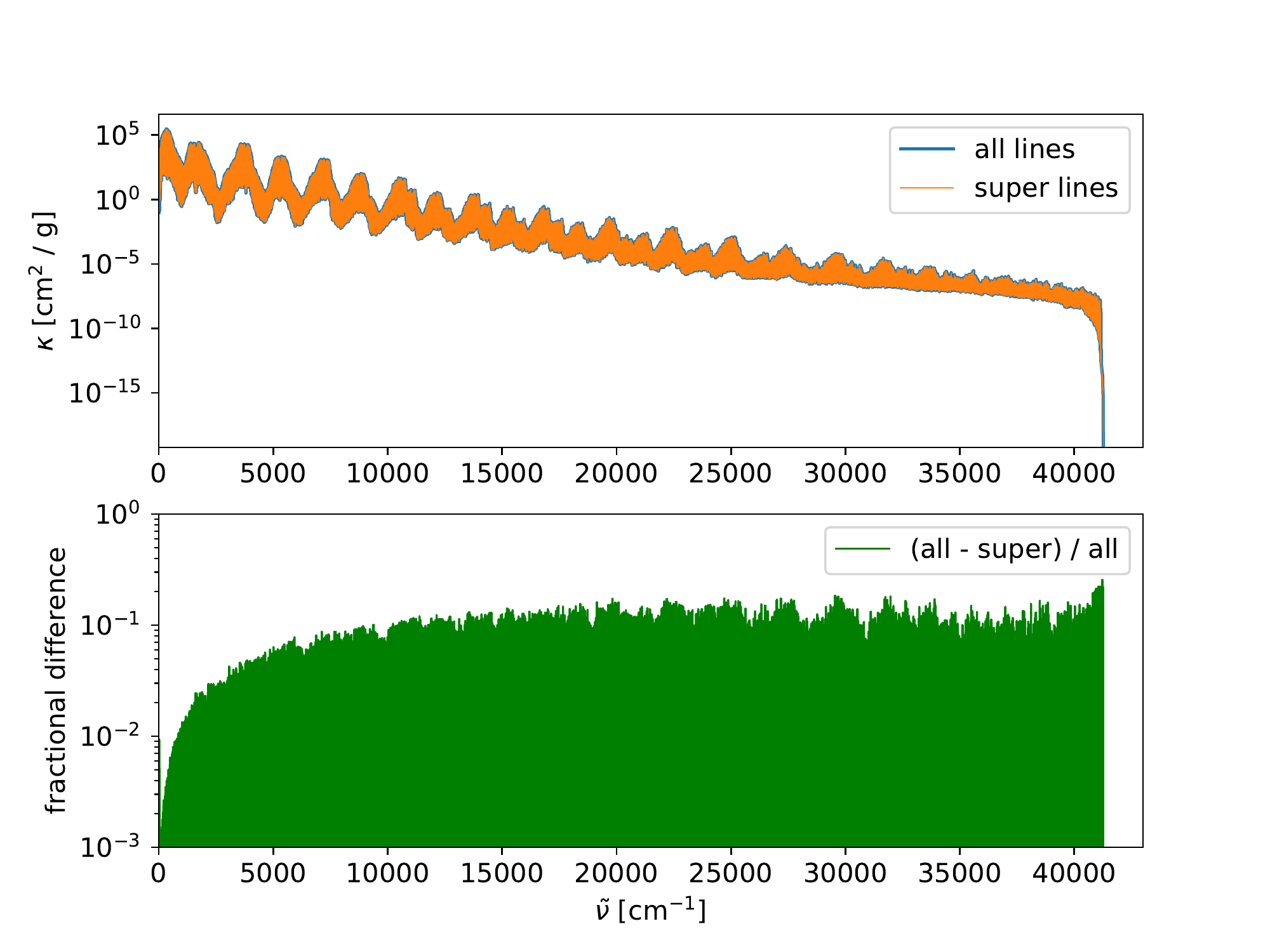}
\end{center}
\caption{Comparison of the ExoMol POKAZATEL water line list with ExoMol superlines, for T = 1000 K, p = 1 bar, and $\Delta \nu = 0.01$ cm$^{-1}$. The fractional difference between the opacity from all lines and the superlines is between 10\% and 20\% for most of the wavenumber range.
}
\label{fig:superLine}
\end{figure}

We compare the opacities computed with superlines to opacities from the full ExoMol line lists. 
Figure \ref{fig:superLine} shows an example of the POKAZATEL water line list. While individual transition lines are not represented in the superlines, the overall opacities agree well, with a fractional difference of less than 20\%. By using superlines, the computational speed of the example in Figure \ref{fig:superLine} is increased by a factor of 720. 

\subsection{The effects of line lists on atmospheric calculations}

In the following section, we estimate the impact on atmospheric modeling when using different line lists. Herein, we compare two different line lists for water: the older BT2 water line list \citep{barber06} and the newer POKAZATEL water line list \citep{ExoMol2018}. Using those, we run models predicting atmospheric temperatures in radiative-convective equilibrium and corresponding planetary emission spectra. We choose two reference cases: (i) a typical hot Jupiter with a primordial H$_2$/He envelope with solar elemental abundances and (ii) a super-Earth with a water-steam atmosphere.

The atmospheric radiative transfer is calculated with the radiative-convective code \texttt{HELIOS} \citep{malik17, malik19a}. Details of the opacity and chemistry calculations are given in \citet{malik19b}. The planetary parameters for HD 189733b and GJ 1132b are chosen as in \cite{malik17, malik19b}. For simplicity, the host star is modeled as a blackbody in both cases.

\begin{figure*}
\begin{center}
\includegraphics[width=2.0\columnwidth]{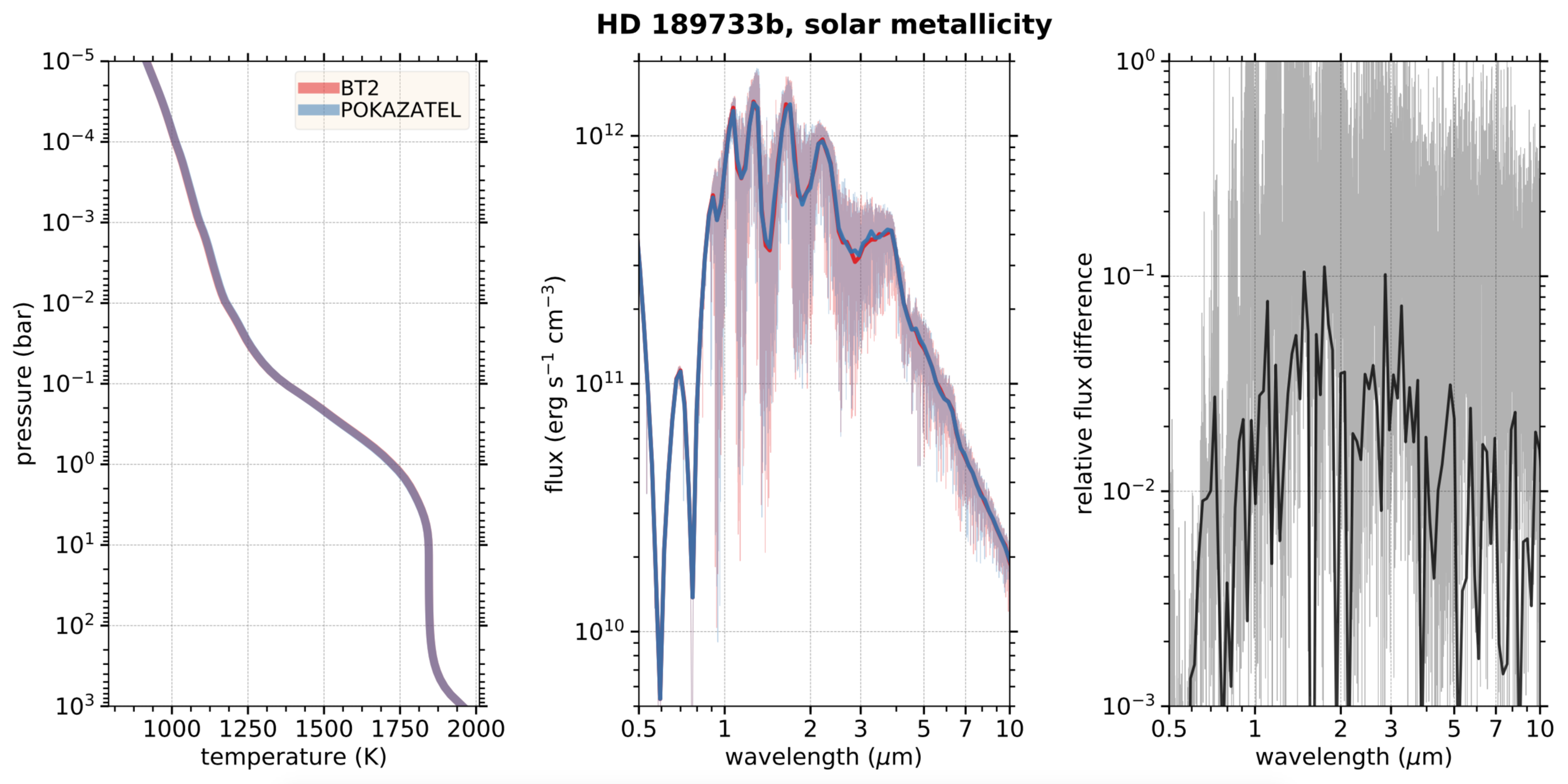}
\includegraphics[width=2.0\columnwidth]{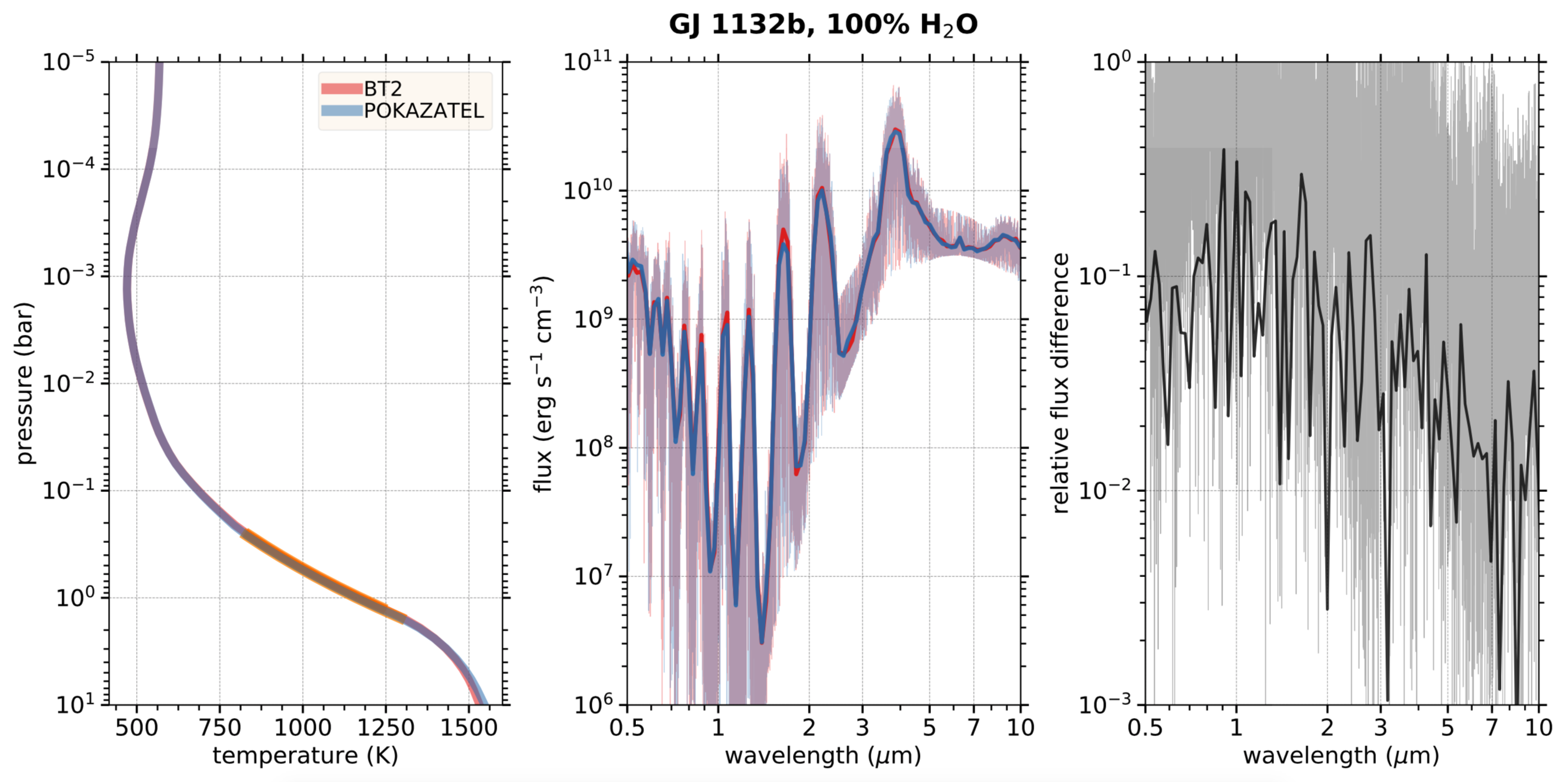}
\end{center}
\caption{TP profiles and emission spectra for a hot Jupiter with a solar abundance atmosphere (top) and a super-Earth with a water-steam atmosphere (bottom), calculated using the BT2 and the POKAZATEL line lists for the water opacities. The right panels show the relative difference in emitted flux between the BT2 and POKAZATEL models. We find relative differences of up to $10 - 40\%$ ($R=30$) and $> 100\%$ ($R=3000$) for certain wavelengths in the near-infrared. These translate to an absolute difference $\sim 30$ ppm ($\sim 2$ ppm) in the secondary eclipse signal of HD 189733b (GJ 1132b) at certain wavelengths in the near-infrared.}
\label{fig:helios_comp}
\end{figure*}

The top row of Figure \ref{fig:helios_comp} shows the results for the hot Jupiter test case in the form of a vertical temperature-pressure (TP) profile, planetary emission spectrum, and relative difference in emitted flux. The latter is calculated as $\vert F_{\rm POK} - F_{\rm BT2}\vert / F_{\rm POK}$, with $F_{\rm POK}$ and $F_{\rm BT2}$ being the spectral fluxes from the POKAZATEL and BT2 models at a given wavelength. The spectra are calculated at the native resolution $R=3000$ and also shown downsampled to $R=30$. Choosing different water line lists causes a negligible effect on the vertical TP profile, with a temperature difference of $< 2$ K for any layer. However, due to the unequal presence and strength of spectral lines between the two line lists, the difference in the planetary emission spectrum is significant. The relative difference is higher than $100\%$ (up to $10\%$) for $R=3000$ (for $R=30$) at certain wavelengths in the near-infrared. The large dependency of this result on the spectral resolution is not surprising considering that the impact of individual lines decreases with decreasing spectral resolution.

The water world case is shown in the bottom row of Figure \ref{fig:helios_comp}. Here the temperatures in the bottom atmosphere deviate between the two models by around 15 K. Consequently, and also because water is the only atmospheric absorber, the relative difference in the spectrum is larger than in the hot Jupiter case, namely, over $30\%$ for some wavelengths at $R=30$. Since POKAZATEL includes more spectral lines than BT2, the emission of the POKAZATEL model is often smaller by the relative amount shown. 

In terms of observational consequences, the absolute difference associated with the secondary eclipse depth is up to $\sim 40$ ppm for the hot Jupiter case and up to $\sim 2$ ppm for the water world case in the near-infrared range of wavelengths.

\subsection{Performance}
\label{performance}
An important part of the opacity calculation is the reading of the line list files. The performance therefore depends highly on the memory access speed. To quantify this dependency, we first test how long it takes to read the preprocessed BT2 water line list files  \citep{barber06} (15 Gbyte, $\tilde\nu$ range 0--30000 cm$^{-1}$) from memory. We test three scenarios: reading the file over a network from a different server, reading from the hard disk, and reading from RAM. We also test three different systems.
\begin{itemize}
    \item A desktop machine (3.6GHz) with a GeForce GTX 980 GPU and ~10 Gbyte of free RAM. This machine does not have enough RAM to store the entire BT2 line list. 
    \item A GPU server (1.7 GHz) with four GeForce GTX 1080 GPUs and 32 Gbyte of RAM.
    \item A cluster node (2.2GHz) on the Ubelix supercomputer with eight GeForce RTX 2080ti GPUs and 256 Gbytes of RAM.
\end{itemize}
For all tests, we use only a single GPU.

Shown in Table \ref{tab:timing0} is the time needed to read the entire BT2 water line list from different memory types. It is clear that only reading the data files can take a substantial amount of time, and it is important to carefully plan memory usage before doing large opacity calculations. Table \ref{tab:timing0} indicates that there can be very large differences associated with the memory speed, depending on the system configuration. 

Shown in Table \ref{tab:timing1} is the time needed for the entire calculation to complete (including memory access) for different computer systems. Again, the measured times clearly show that the performance depends highly on the memory speed. If the line list fits into the system RAM, then one can achieve a significantly higher performance than reading from a hard disk or even over a network cable. 

A way to further increase the performance is to combine opacity calculations from different pressure values into a single run. In this way, the number of data reads can be reduced because the data remain on the GPU for the different pressure point calculations. This effect is shown in Table \ref{tab:timing2}. By using 10 pressure points in a single run, the overall performance of low-resolution runs can be improved by about 60--70\%. In high-resolution runs, the effect is not as strong. 

\begin{figure}[ht]
\begin{center}
\includegraphics[width=1.0\columnwidth]{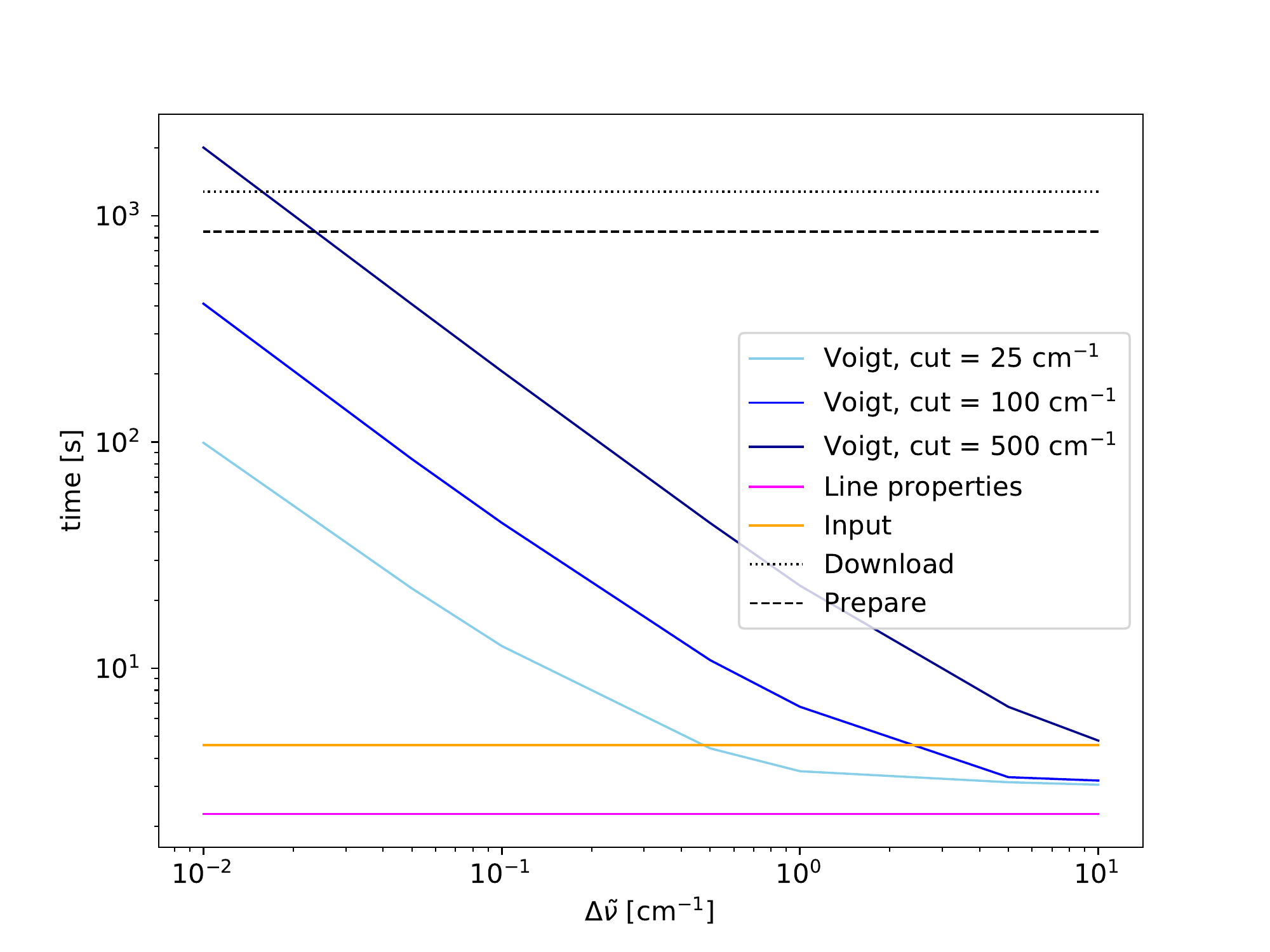}
\end{center}
\caption{Timing of different parts of \texttt{HELIOS-K} for the example of the BT2 water line list (0 - 30000 cm$^{-1}$). The Voigt function calculations are in dark blue, blue, and light blue for the cutting lengths of 500, 100, and 25 cm$^{-1}$. Reading the line list (from RAM) is in orange, and preparing the transition line properties is in magenta. Also shown is the time needed to download and unpack the line list from the ExoMol servers and to preprocess it into the \texttt{HELIOS-K} binary format. For low-resolution calculations, the reading part of the line list dominates the opacity calculation. Timings are measured with an Nvidia GTX 1080 GPU.}
\label{fig:timing}
\end{figure}

Shown in Figure \ref{fig:timing} are timing measurements for different parts of \texttt{HELIOS-K} for the example of the BT2 water line list. One can see that for low-resolution opacity calculations, the reading part of the line list dominates the entire calculation. The Voigt calculation depends on the resolution and cutting length, while the other parts depend only on the number of transition lines in the line list database. Also shown is the time needed to download the line lists from the ExoMol website and to preprocess the line list into the \texttt{HELIOS-K} binary file format. These two timings depend on the used internet connection speed and the CPU clock rate of the used machine.

\texttt{HELIOS-K} can perform a typical opacity calculation with a resolution of $\Delta \tilde{\nu} = 0.01$ with $10^7$ lines in about 1 s.

For comparison, \cite{Yurchenko+2018} listed the timing of the \texttt{ExoCross} code as 251 s, excluding the data reading and by using approximations for the line wings. \texttt{HELIOS-K} performs the entire calculation in 12.5 s, including the data reading time.  The traditional Humlicek approach takes 2775.6 s (excluding the data reading time).

\begin{table}[h]
\centering
\begin{tabular}{c|c|c|c|c}

 System & GPU & $\tilde\nu$ Range in cm$^{-1}$ & Data Location & Time  \\
  \hline
  \hline 
 Desktop & GTX 980 & 0 - 30000 & Network & 252 s \\
 Desktop & GTX 980 & 0 - 30000 & Hard disk & 30.0 s \\
 Desktop & GTX 980 & 0 - 30000 & Partially RAM & 12.5 s \\
 Server & GTX 1080 & 0 - 30000 & Hard disk & 40.5 s \\
 Server & GTX 1080 & 0 - 30000 & RAM & \textbf{1.9} s \\
 Cluster & RTX 2080ti & 0 - 30000 & Hard disk & 55.5 s \\
 Cluster & RTX 2080ti & 0 - 30000 & RAM & 4.8 s \\
\end{tabular}
\caption{Time to read the entire BT2 water line list for different machines and memory types. The GPUs are not involved in this test, but are listed for a better overview of the system types. The measured times indicate that the memory type used can highly affect the performance.}
\label{tab:timing0}
\end{table}

\begin{table}[h]
\centering
\begin{tabular}{c|c|c|c}

 GPU & $\tilde\nu$ Range in cm$^{-1}$ & Data Location & Time  \\
  \hline
  \hline 
 GTX 980 & 0 - 5000 & Over network & 172.6 s \\
 GTX 980 & 0 - 5000 & Hard disk & 20.9 s \\
 GTX 1080 & 0 - 5000 & Hard disk & 26.3 s \\
 GTX 980 & 0 - 5000 & RAM & 13.9 s \\
 GTX 1080 & 0 - 5000 & RAM &  \textbf{8.8 s} \\
  \hline
 GTX 980 & 0 - 30000 & Partially in RAM & 30.7 s \\
GTX 1080 & 0 - 30000 & RAM &  \textbf{16.5 s} \\
\end{tabular}
\caption{Performance of BT2 water line list for $\Delta \tilde{\nu} = 0.1$ cm$^{-1}$ and a cutting length of 25 cm$^{-1}$. An essential factor of the overall performance is the memory access of the line lists. For the best performance, the data must remain fully in RAM.}
\label{tab:timing1}
\end{table}

\begin{table}[h]
\centering
\begin{tabular}{c|c|c|c|c}

 GPU & nP & $\tilde{\nu}$ Range in cm$^{-1}$  & $\Delta \tilde{\nu}$ in cm$^{-1}$  & Time  \\
  \hline
  \hline 
 GTX 1080 & 1 & 0 - 30000 & 10.0 & 10.1 s \\
 GTX 1080 & 1 & 0 - 30000 & 1.0  & 10.4 s \\
 GTX 1080 & 1 & 0 - 30000 & 0.1  & 16.5 s \\
 GTX 1080 & 1 & 0 - 30000 & 0.01 & 99.2 s \\
 \hline
 RTX 2080ti & 1 & 0 - 30000 & 10.0 & 10.7 s \\
 RTX 2080ti & 1 & 0 - 30000 & 1.0  & 10.8 s \\
 RTX 2080ti & 1 & 0 - 30000 & 0.1  & 12.5 s \\
 RTX 2080ti & 1 & 0 - 30000 & 0.01 & 44.7 s \\
 \hline
 GTX 1080 & 10 & 0 - 30000 & 10.0 & 66 s \\
 GTX 1080 & 10 & 0 - 30000 & 1.0  & 76 s \\
 GTX 1080 & 10 & 0 - 30000 & 0.1  & 165 s \\
 GTX 1080 & 10 & 0 - 30000 & 0.01 & 1044 s \\
 \hline
 RTX 2080ti & 10 & 0 - 30000 & 10.0 & 57 s \\
 RTX 2080ti & 10 & 0 - 30000 & 1.0  & 66 s \\
 RTX 2080ti & 10 & 0 - 30000 & 0.1  & 99 s \\
 RTX 2080ti & 10 & 0 - 30000 & 0.01 & 475 s \\
\end{tabular}
\caption{Performance of BT2 water line list for $\tilde{\nu}$ = 0--5000 cm$^{-1}$, cutting length 25cm$^{-1}$. nP indicates the number of pressure points in the calculation. At low resolution, using multiple pressure points can improve the performance.}
\label{tab:timing2}
\end{table}

\section{Discussion}

We  present an improved and more powerful version of the GPU opacity calculator \texttt{HELIOS-K}. For certain species, the code achieves a speed-up of nearly 2 orders of magnitude since the first version presented in \cite{gh15}. The code also supports more features and allows a simple use of different line list databases. During our work, we encountered several issues with multiple databases, which we reported to the maintainers. Our new code is fast enough to process the current largest molecular line lists from ExoMol. 

While also the progress in the speed of modern GPUs helped to reduce the computing time, the largest increase in speed comes from the new parallelization algorithm, especially the usage of asynchronous and simultaneous data read and computation techniques, which turn out to be very powerful. It is not only the GPU itself that makes the speed-up, but also the efficient interaction between CPU, GPU, and memory as a full system.

But it is clear that when even bigger molecular line lists get published, new challenges could occur in both computing time and storage capacity. Atomic opacities are not as challenging in terms of computing power, but there we are faced with the incompleteness of theory. When atomic opacities are approximated by Voigt profiles with Lorentzian or sub-Lorentzian line wings, it can introduce significant errors in the opacity function. This is especially true for resonance and autoionization lines, where broadening mechanisms are not known well enough. 

We computed opacity functions for several hundred species, which we store in our database and share with the community via the website www.opacity.world and our data server https://chaldene.unibe.ch or via https://dace.unige.ch/opacity and https://dace.unige.ch/opacityDatabase.

It is evident that opacity calculations are a complex, tedious undertaking.  In order to ensure reproducibility, the computed opacities have been made freely available on the Swiss PlanetS platform DACE (\texttt{https://dace.unige.ch}).  To aid the user in developing intuition, the opacities may be visualized via a graphical user interface (\texttt{https://dace.unige.ch/opacity}).  Opacities may be downloaded at the full spectral resolution (0.01 cm$^{-1}$) in binary format (\texttt{https://dace.unige.ch/opacityDatabase}).  Alternatively, the user may specify customized arrays of temperature and pressure, as well as the range of wavenumbers or wavelengths required, and interpolated opacities will be produced in the HDF5 format. In the future, we envision a ``versioning" capability, where different implementations using the same generation of spectroscopic line lists (e.g., by different researchers) or implementations using different generations of line lists may be archived, so that differences in the subsequent models or simulations may be diagnosed.  It is our belief that these capabilities are critical for the exoplanet community as we move into the era of precision spectroscopy of exoplanetary atmospheres with the James Webb Space Telescope and the next generation of large ground-based telescopes.

\begin{acknowledgments}
Calculations were performed on UBELIX (http://www.id.unibe.ch/hpc), the HPC cluster at the University of Bern. This work has made use of the VALD database, operated at Uppsala University, the Institute of Astronomy RAS in Moscow, and the University of Vienna. We acknowledge partial financial support from the Swiss National Science Foundation, the European Research Council (via a Consolidator Grant to K.H.; grant No. 771620), the PlanetS National Center of Competence in Research (NCCR), and the Center for Space and Habitability (CSH), and the Swiss-based MERAC Foundation.
\end{acknowledgments}

\newpage

\bibliography{mybib}

\appendix
\section{Atomic Opacities}
The following Figures (\ref{fig:atomic0} - \ref{fig:atomic3}), together with Figure \ref{fig:atomic} in the main text, give an overview of 48 atomic opacities, for the NIST, Kurucz, and VALD3 databases.

\label{appendixA}
\begin{figure*}[h]
\begin{center}
\includegraphics[width=1.0\columnwidth]{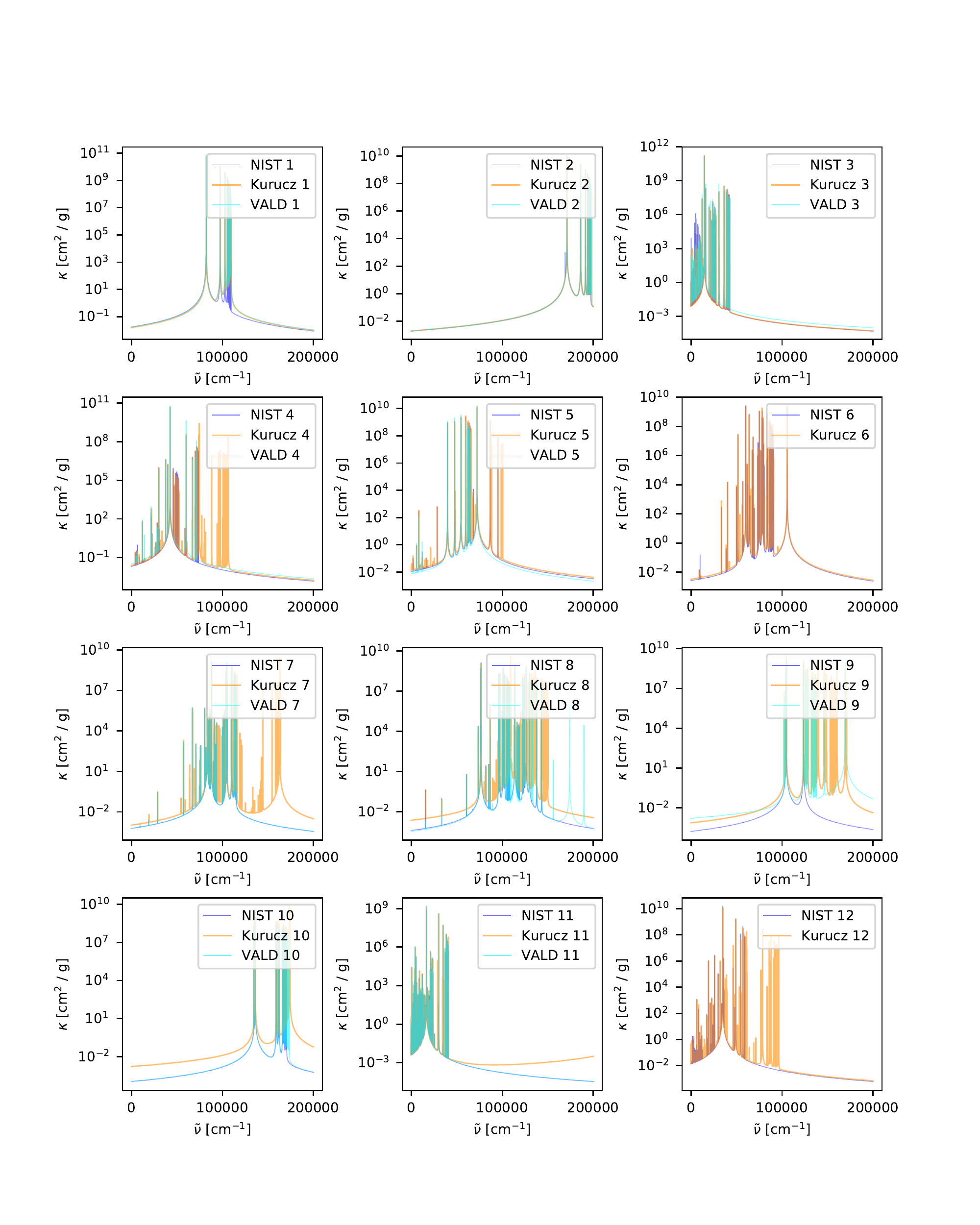}
\end{center}
\caption{Comparison of atomic opacities between the NIST, Kurucz, and VALD3 databases, elements 1 to 12. We use a temperature of 3000 K and no cutting length.}
\label{fig:atomic0}
\end{figure*}

\begin{figure*}[h]
\begin{center}
\includegraphics[width=1.0\columnwidth]{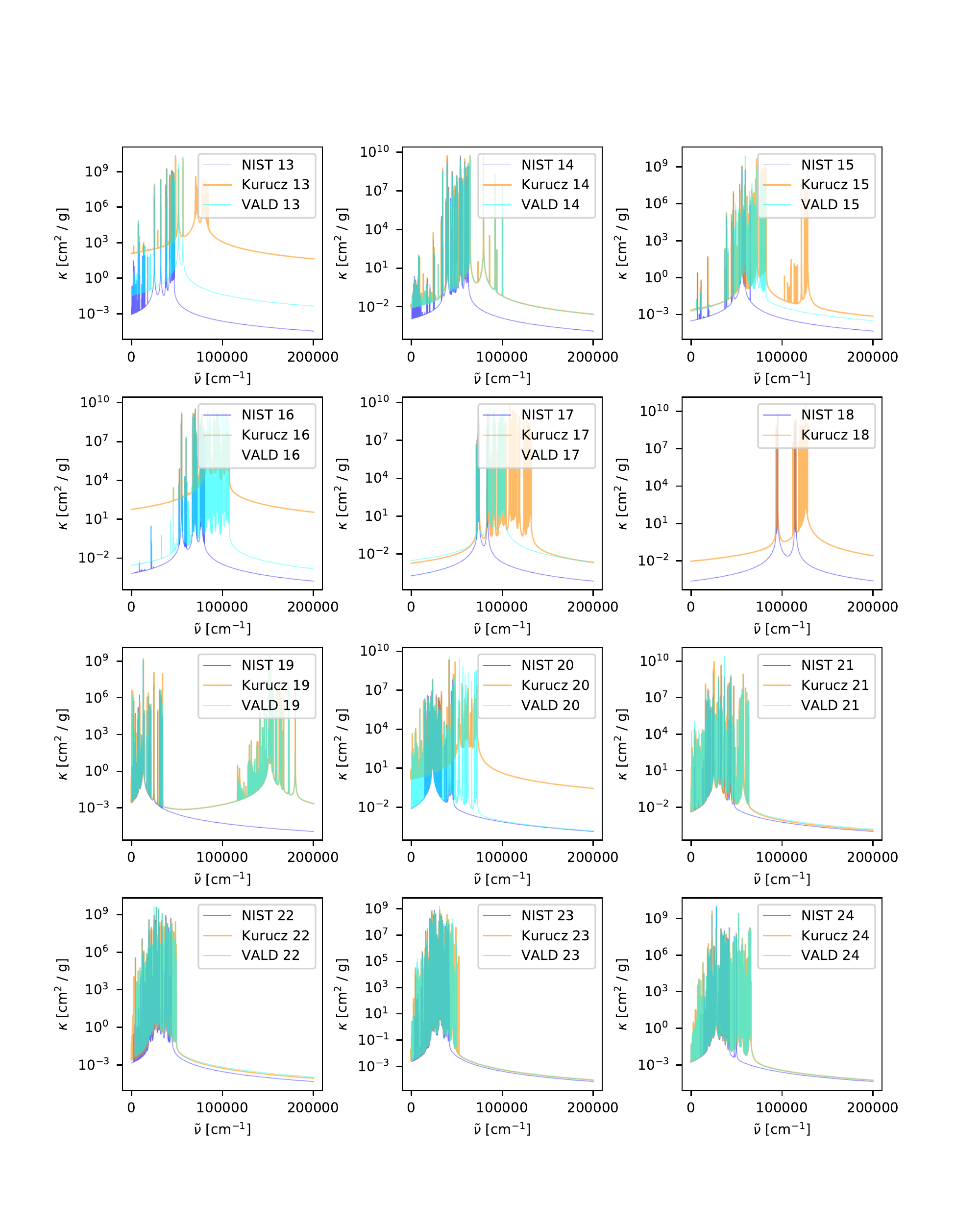}
\end{center}
\caption{Comparison of atomic opacities between the NIST, Kurucz, and VALD3 databases, elements 13 to 24. We use a temperature of 3000 K and no cutting length.}
\label{fig:atomic1}
\end{figure*}

\begin{figure*}[h]
\begin{center}
\includegraphics[width=1.0\columnwidth]{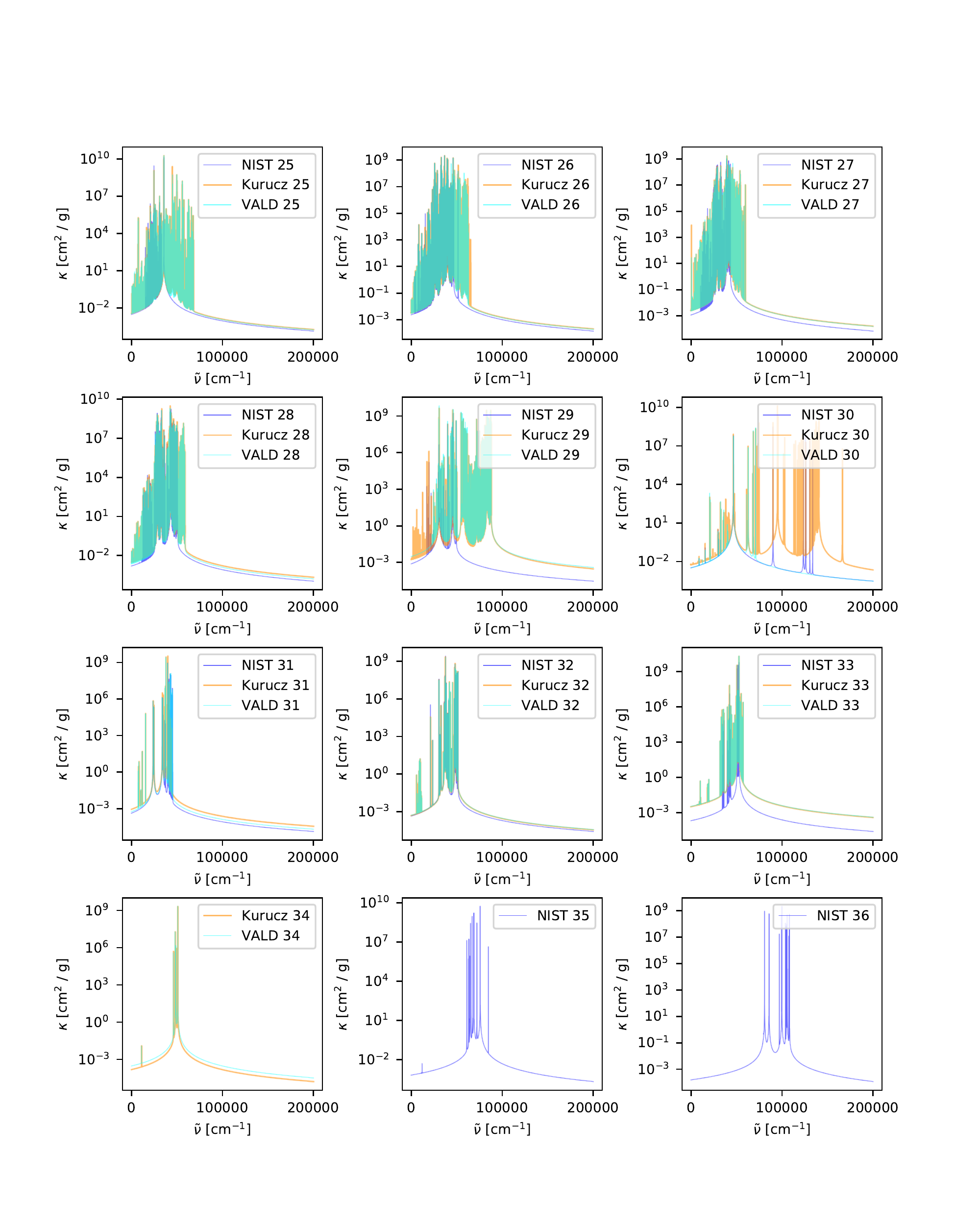}
\end{center}
\caption{Comparison of atomic opacities between the NIST, Kurucz, and VALD3 databases, elements 25 to 36. We use a temperature of 3000 K and no cutting length.}
\label{fig:atomic2}
\end{figure*}

\begin{figure*}[h]
\begin{center}
\includegraphics[width=1.0\columnwidth]{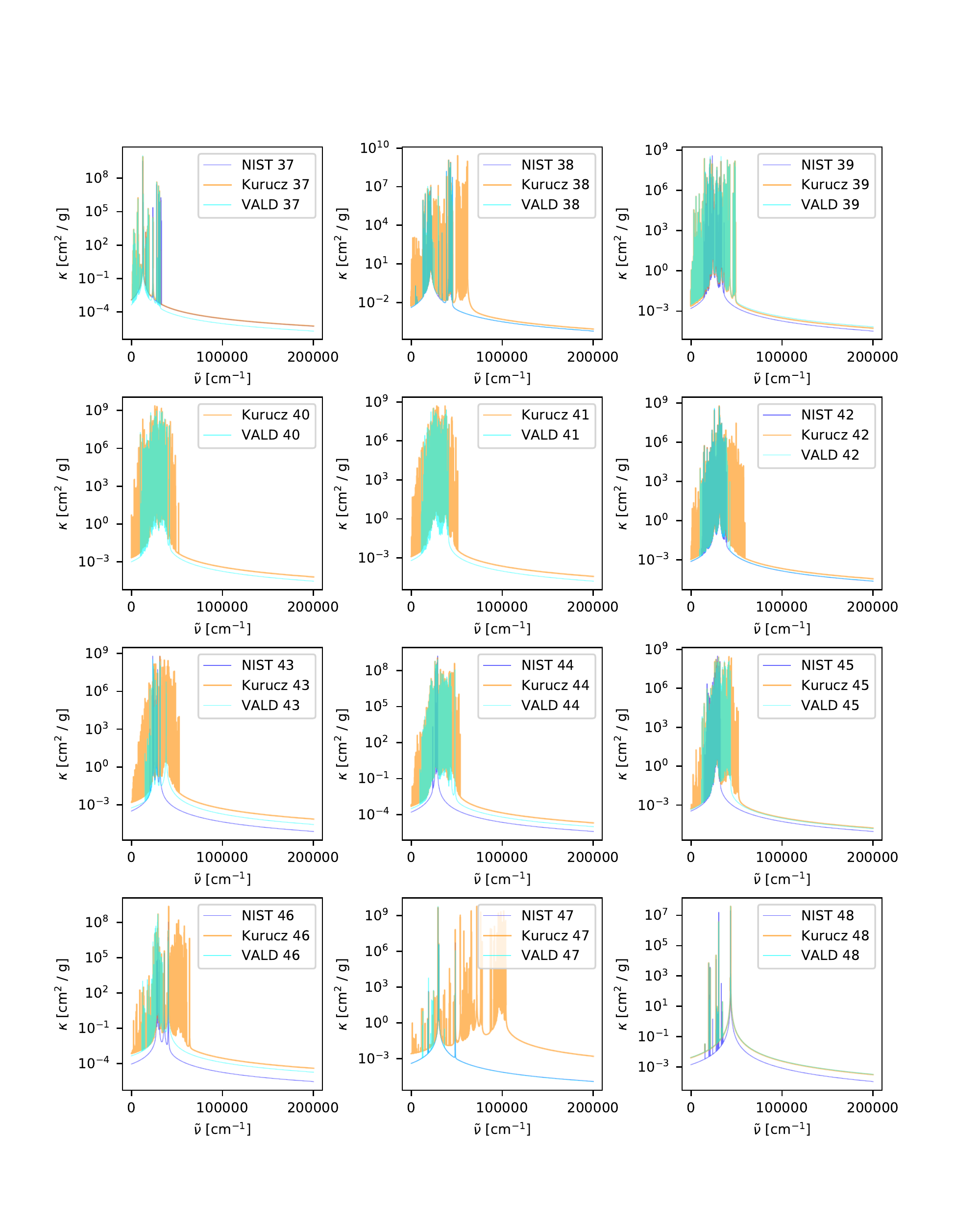}
\end{center}
\caption{Comparison of atomic opacities between the NIST, Kurucz, and VALD3 databases, elements 37 to 48. We use a temperature of 3000 K and no cutting length.}
\label{fig:atomic3}
\end{figure*}

\newpage
\section{Molecular Opacity Comparison}
\label{appendixB}

The following Figures (\ref{fig:C0} - \ref{fig:C2}) show a comparison between \texttt{HELIOS-K} cross sections and data from the ExoMol website.  

\begin{figure*}[h]
\begin{center}
\includegraphics[width=0.9\columnwidth]{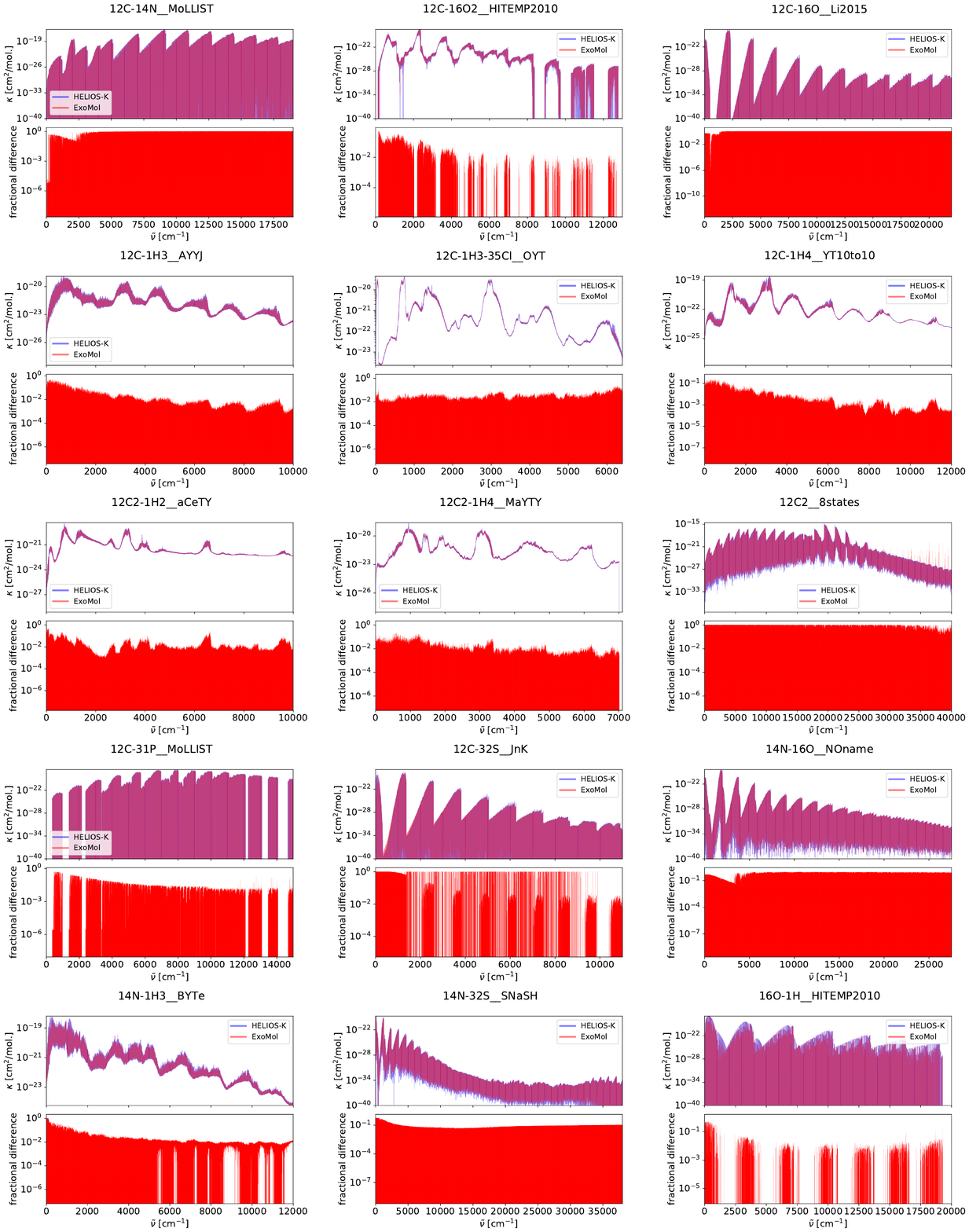}
\end{center}
\caption{Comparison between cross sections calculated with \texttt{HELIOS-K} and from the ExoMol website. We use a wavenumber resolution of $\Delta \tilde{\nu} = 0.1$ cm$^{-1}$ and T=1500 K (1000 K for 12C2-1H4\_\_MaYTY). }
\label{fig:C0}

\end{figure*}
\begin{figure*}[h]
\begin{center}
\includegraphics[width=0.9\columnwidth]{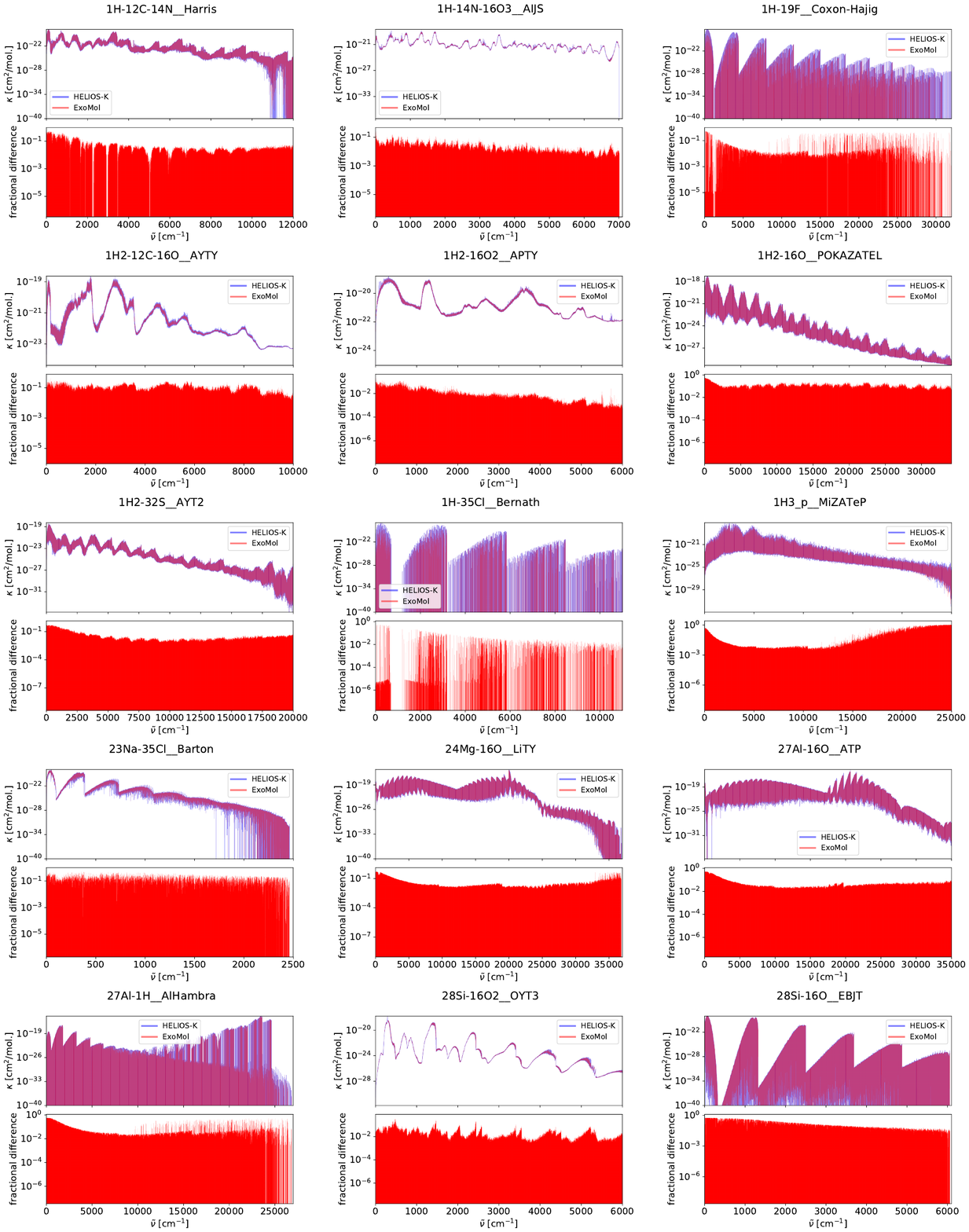}
\end{center}
\caption{Comparison between cross sections calculated with \texttt{HELIOS-K} and from the ExoMol website. We use a wavenumber resolution of $\Delta \tilde{\nu} = 0.1$ cm$^{-1}$ and T=1500 K (500 K for 1H-14N-16O3\_\_AIJS).}
\label{fig:C1}

\end{figure*}
\begin{figure*}[h]
\begin{center}
\includegraphics[width=0.9\columnwidth]{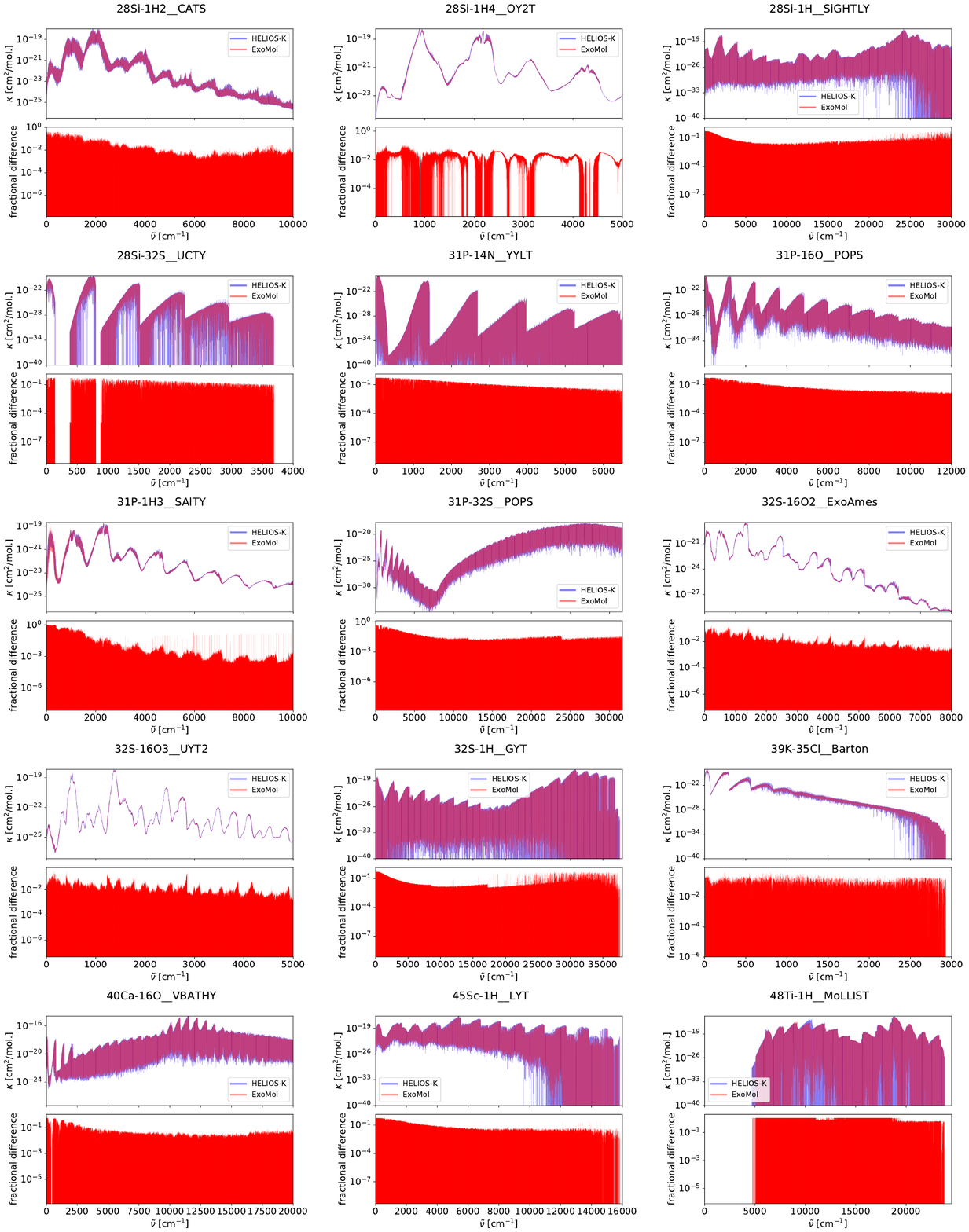}
\end{center}
\caption{Comparison between cross sections calculated with \texttt{HELIOS-K} and from the ExoMol website. We use a wavenumber resolution of $\Delta \tilde{\nu} = 0.1$ cm$^{-1}$ and T=1500 K (1000 K for 32S-16O3\_\_UYT2).}
\label{fig:C2}
\end{figure*}

\newpage
\section{Molecular Opacity Overview}

The following Figures (\ref{fig:fa} - \ref{fig:fd}) show opacities for 30 commonly used molecules.

\begin{figure*}[h]
\begin{center}
\includegraphics[width=1.0\columnwidth]{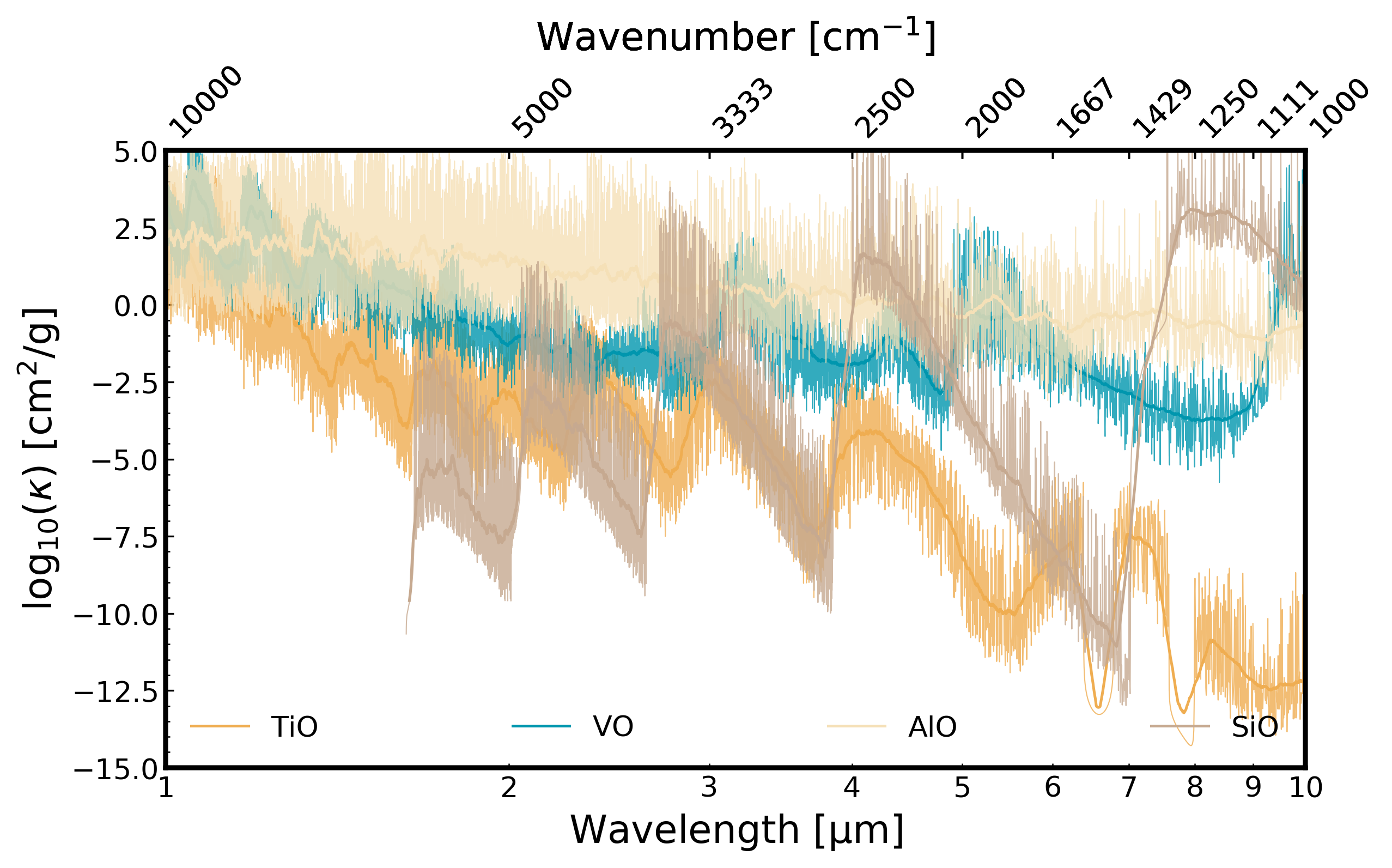}
\includegraphics[width=1.0\columnwidth]{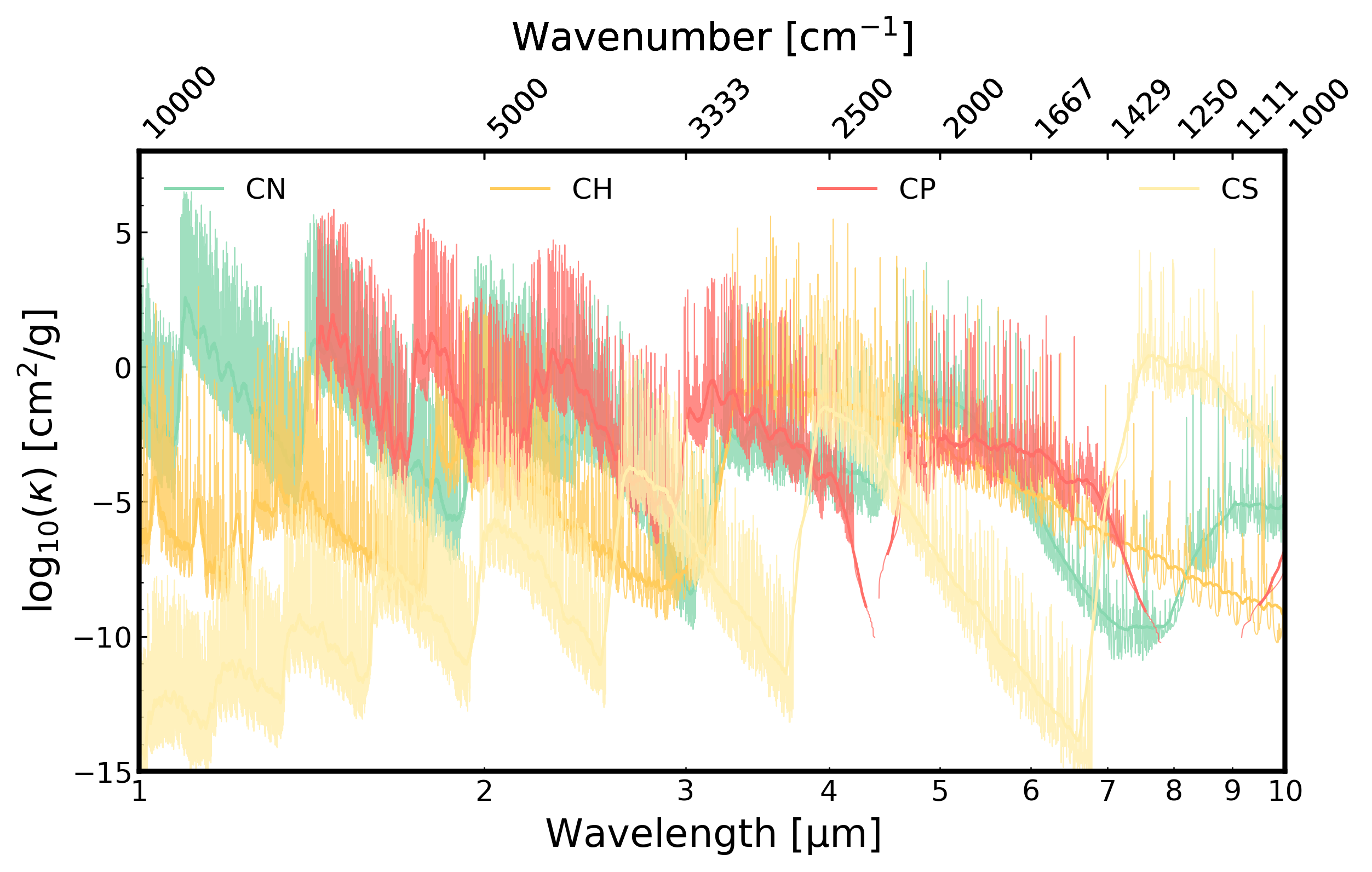}
\end{center}
\caption{Examples of molecular opacities: TiO \citep{ExoMol_TiO}, VO \citep{ExoMol_VO}, AlO \citep{ExoMol_AlO}, SiO \citep{ExoMol_SiO}, CN \citep{ExoMol_CN}, CH \citep{ExoMol_CH}, CP \citep{ExoMol_CP} and CS \citep{ExoMol_CS}. We use a temperature of 1500 K, a pressure of 0.001 bar, and cutting length of 100 cm$^{-1}$.}
\label{fig:fa}
\end{figure*}

\begin{figure*}[h]
\begin{center}
\includegraphics[width=1.0\columnwidth]{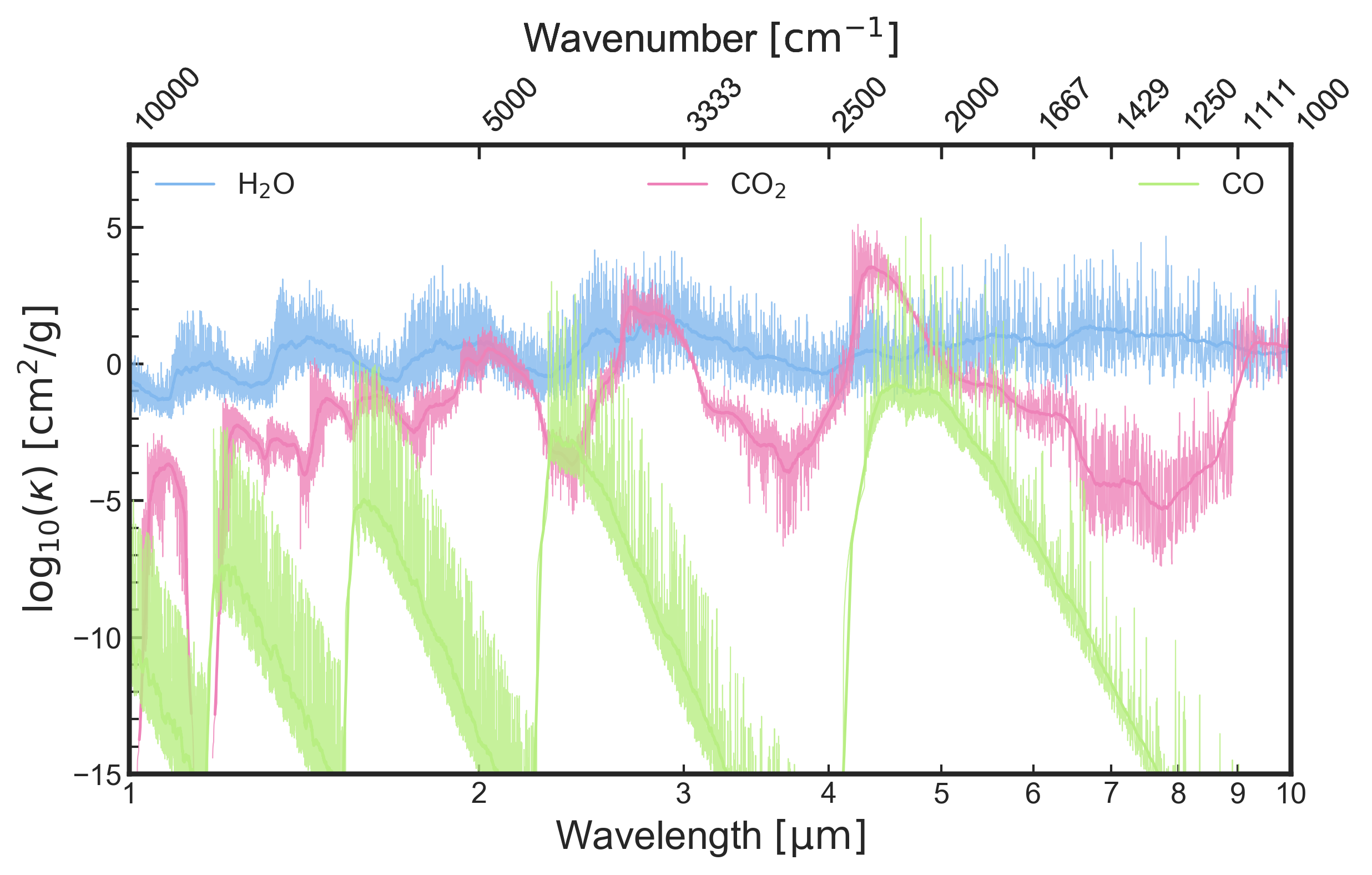}
\includegraphics[width=1.0\columnwidth]{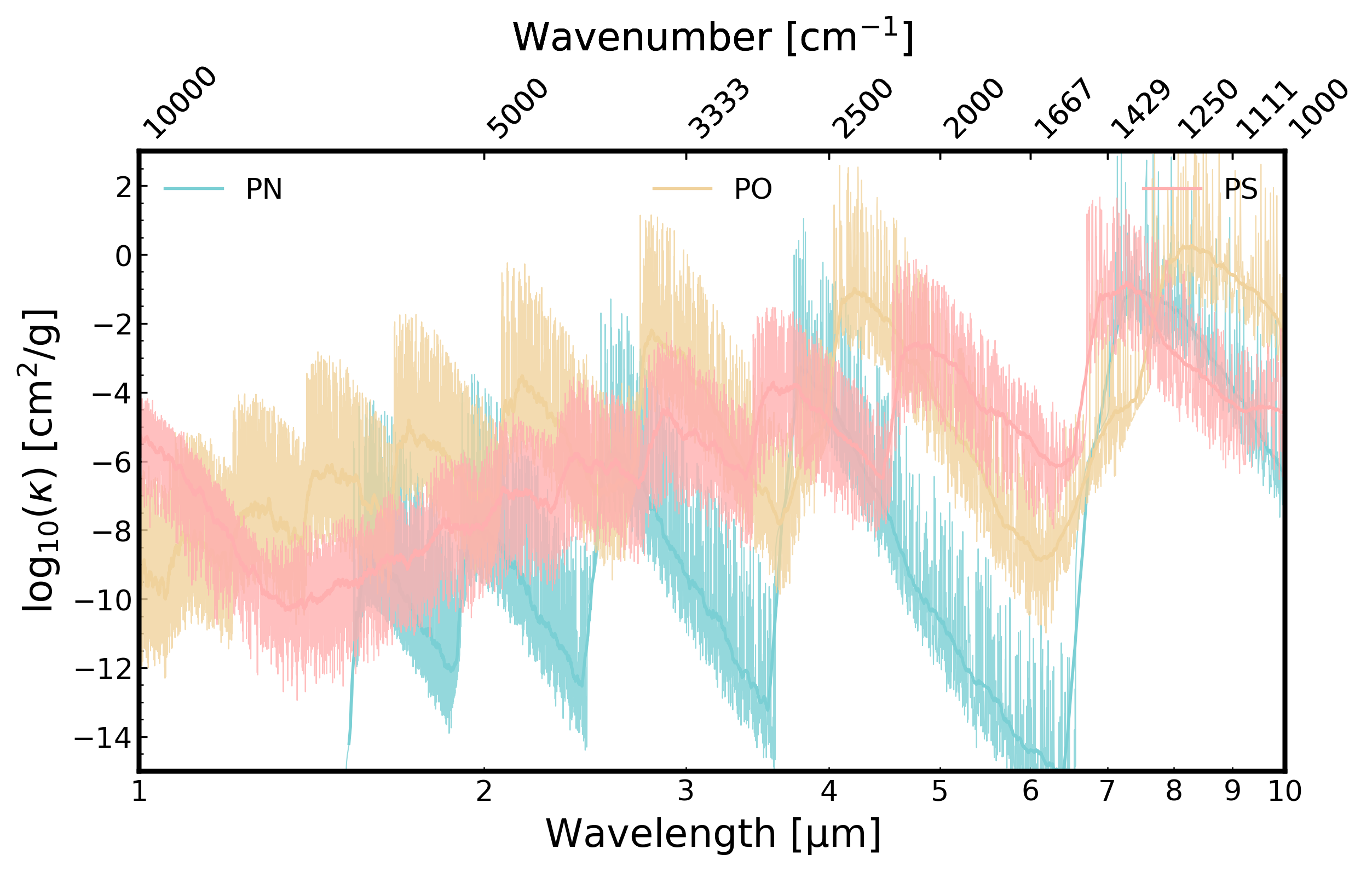}
\end{center}
\caption{Examples of molecular opacities: H$_2$O \citep{ExoMol2018}, CO$_2$ \citep{HITEMP}, CO \citep{ExoMol_CO}, PN \citep{ExoMol_PN}, PO \citep{ExoMol_PO_PS}, and PS \citep{ExoMol_PO_PS}. We use a temperature of 1500 K, a pressure of 0.001 bar, and cutting length of 100 cm$^{-1}$.}
\label{fig:fb}
\end{figure*}

\begin{figure*}[h]
\begin{center}
\includegraphics[width=1.0\columnwidth]{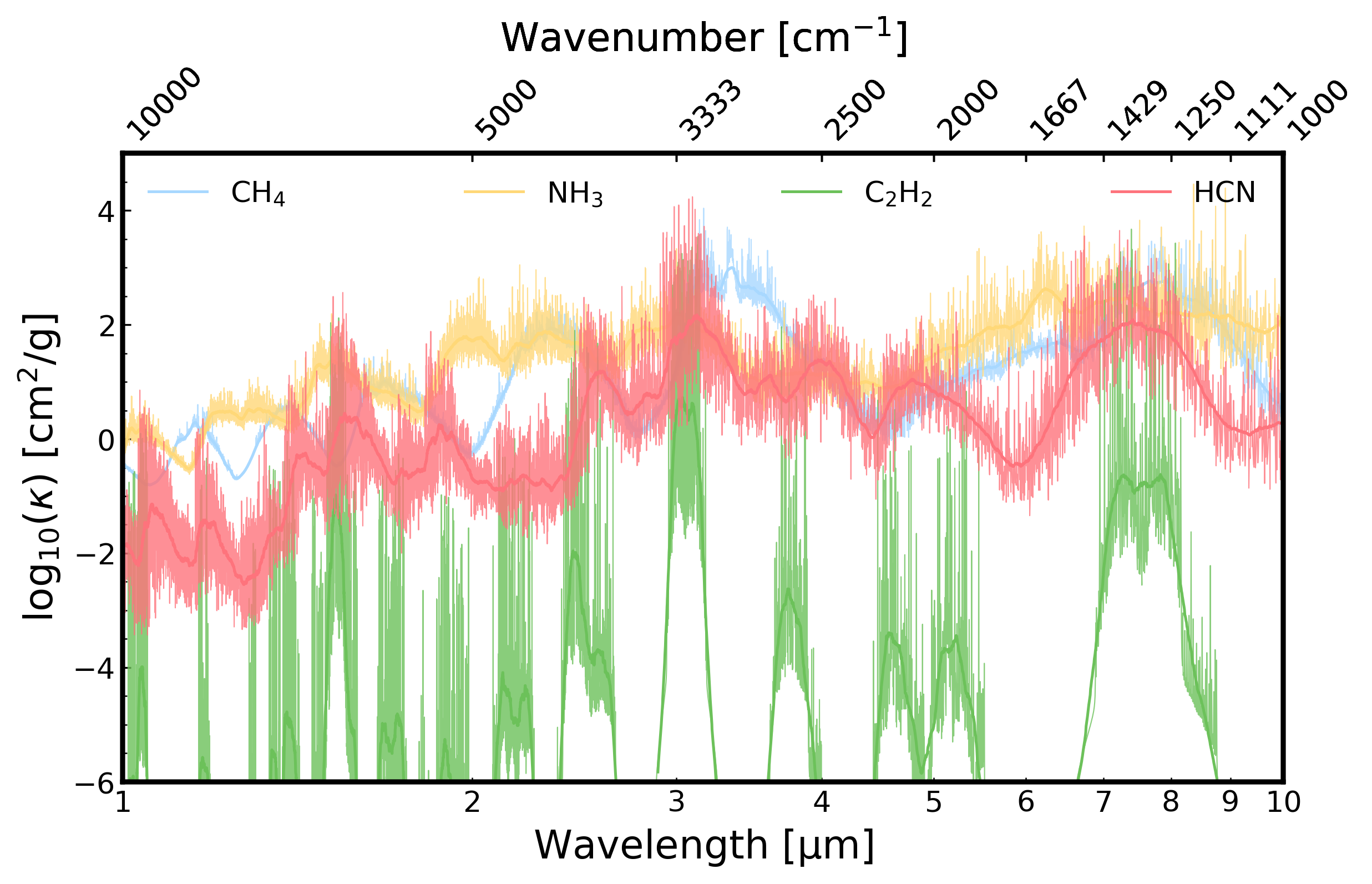}
\includegraphics[width=1.0\columnwidth]{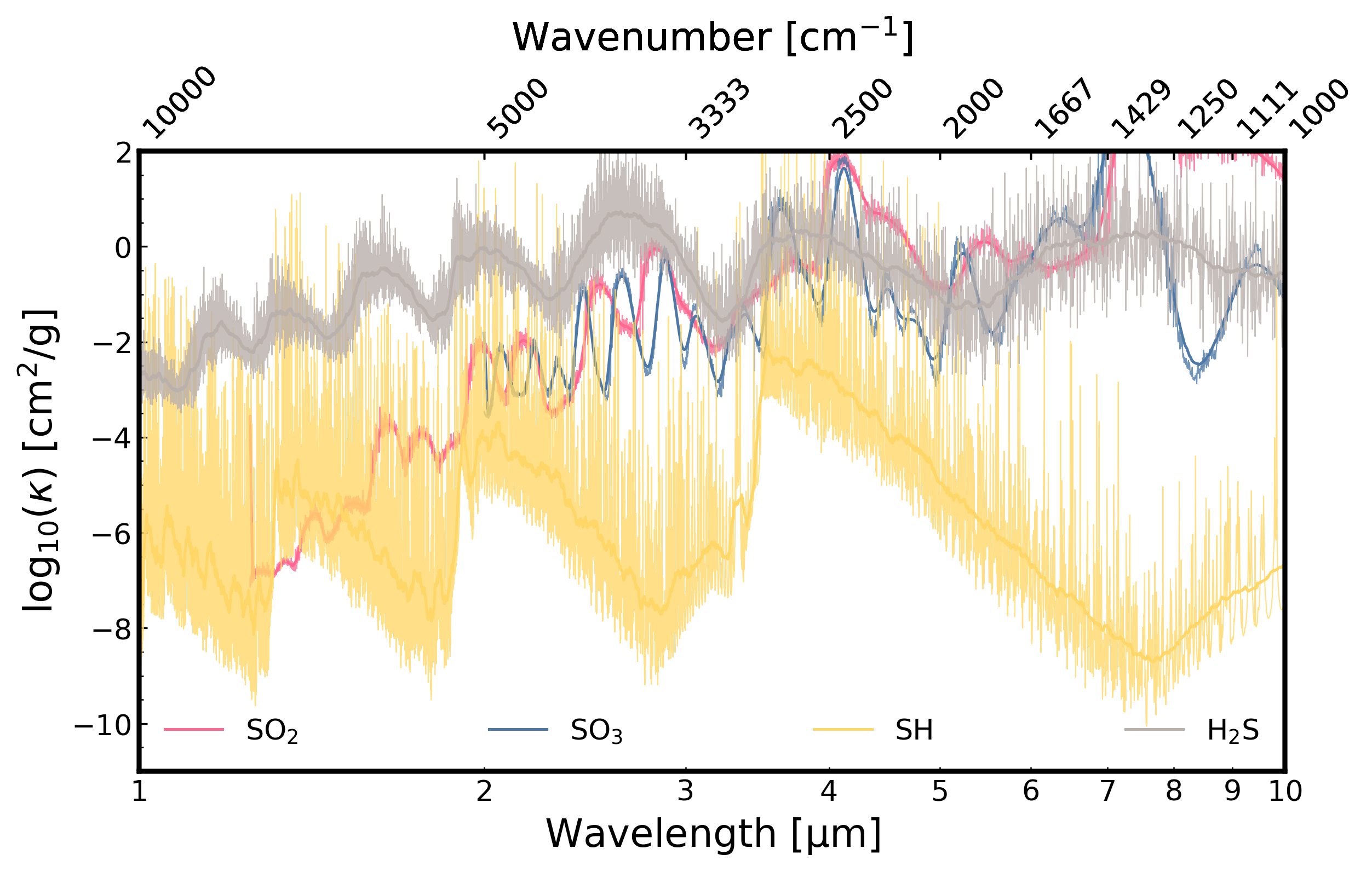}
\end{center}
\caption{Examples of molecular opacities: CH$_4$ \citep{ExoMol_CH4}, NH$_3$ \citep{ExoMol_NH3}, C$_2$H$_2$\citep{ExoMol_C2H2}, HCN \citep{ExoMol_HCN_a, ExoMol_HCN_b}, SO$_2$ \citep{ExoMol_SO2}, SO$_3$ \citep{ExoMol_SO3}, SH \citep{ExoMol_NS_SH}, and H$_2$S \citep{ExoMol_H2S}. We use a temperature of 1500 K (500 K for SO$_3$), a pressure of 0.001 bar, and cutting length of 100 cm$^{-1}$.}
\label{fig:fc}
\end{figure*}

\begin{figure*}[h]
\begin{center}
\includegraphics[width=1.0\columnwidth]{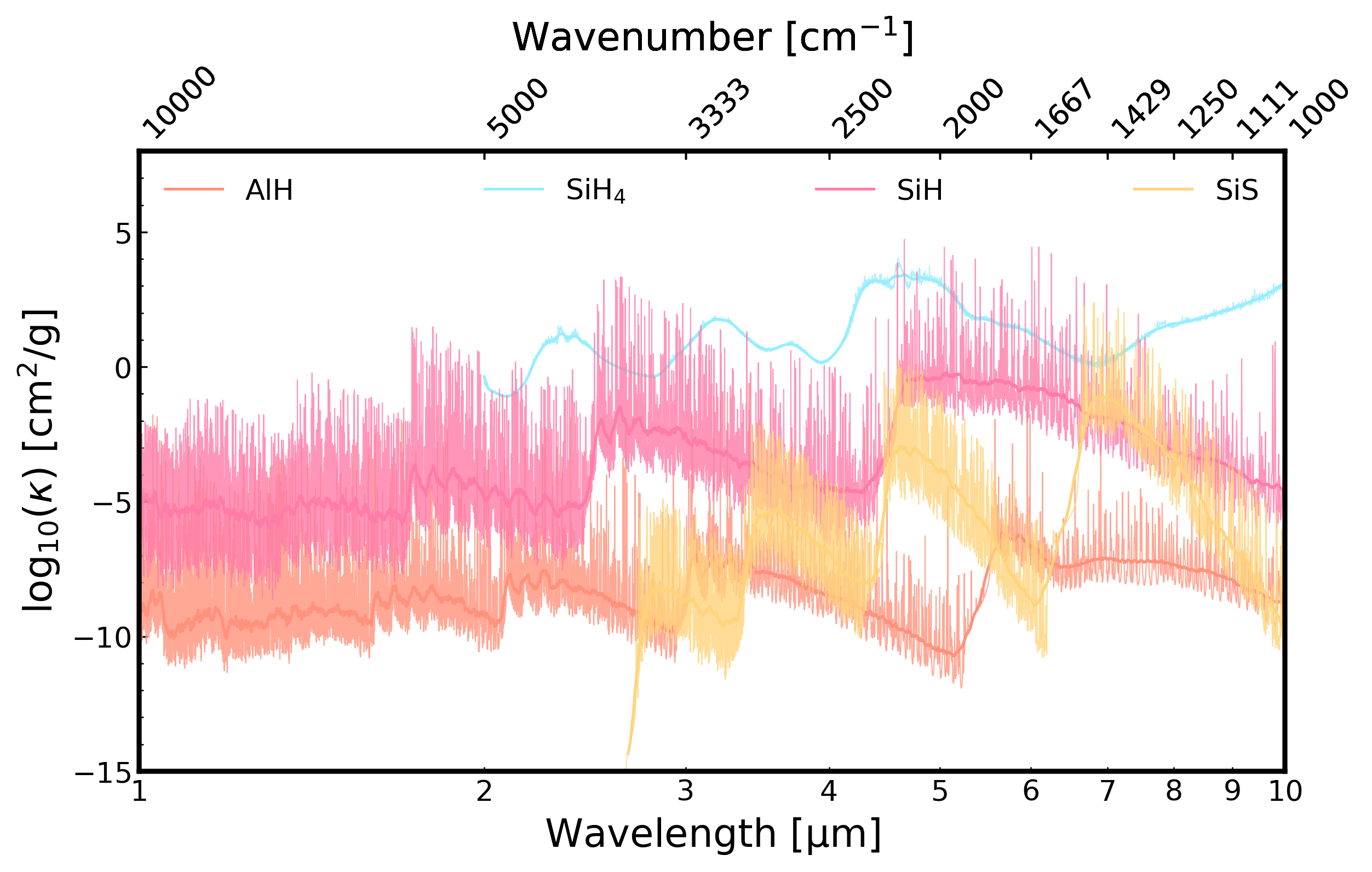}
\includegraphics[width=1.0\columnwidth]{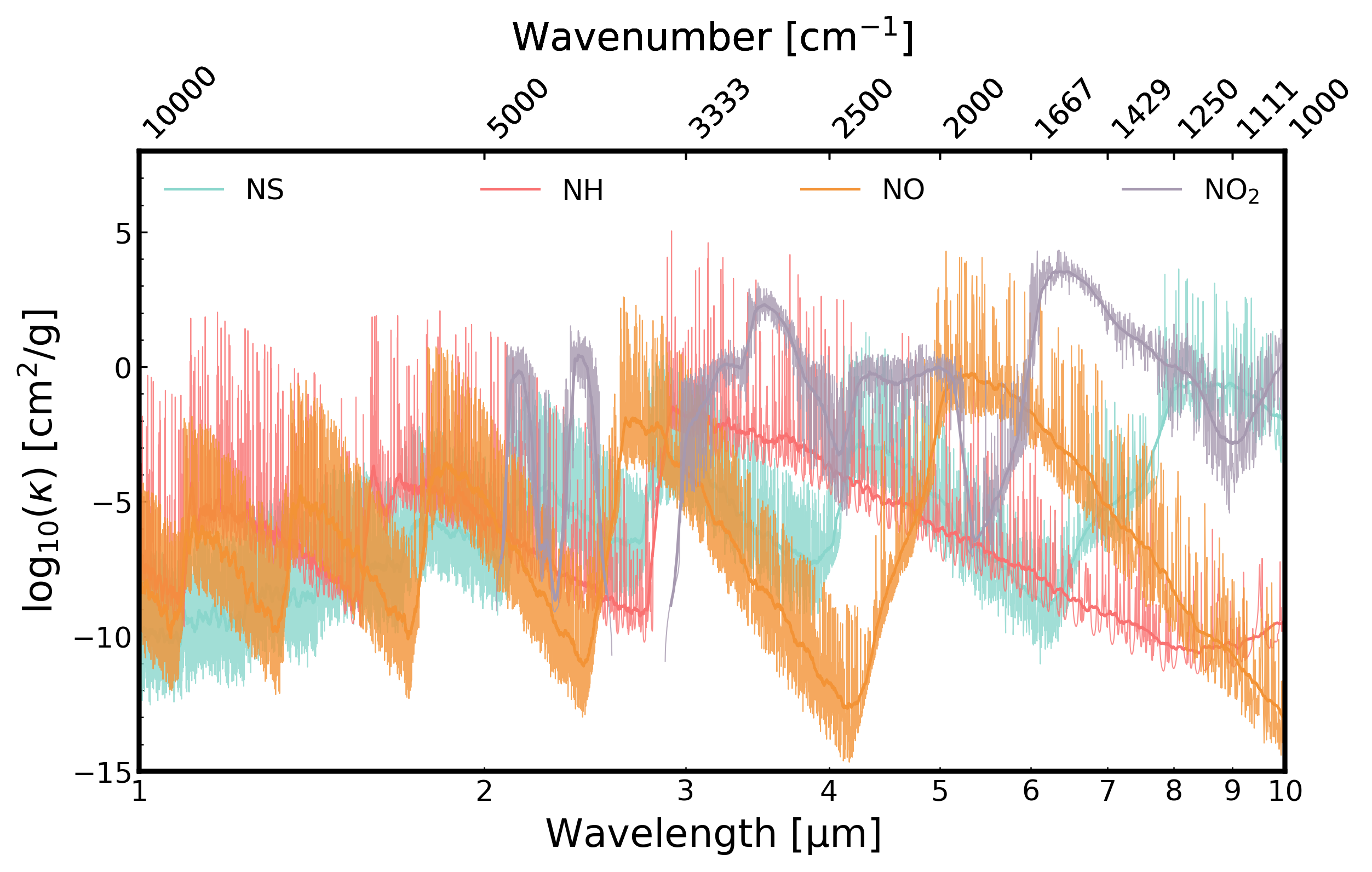}
\end{center}
\caption{Examples of molecular opacities: AlH \citep{ExoMol_AlH}, SiH$_4$ \citep{ExoMol_SiH4}, SiH \citep{ExoMol_SiH}, SiS \citep{ExoMol_SiS}, NS \citep{ExoMol_NS_SH}, NH \citep{ExoMol_NH}, NO \citep{ExoMol_NO}, and NO$_2$ \citep{HITEMP, HITEMP2019}. We use a temperature of 1500 K, a pressure of 0.001 bar, and cutting length of 100 cm$^{-1}$.}
\label{fig:fd}
\end{figure*}

\label{lastpage}

\end{document}